\documentclass[final,5p,times,twocolumn]{elsarticle} 
\DeclareMathAlphabet{\mathpzc}{OT1}{pzc}{m}{it}
\usepackage{lineno}
\usepackage{setspace}
\usepackage{graphicx}
\usepackage{amssymb}
\usepackage{amsmath}
\usepackage{amsthm}
\usepackage[colorlinks=true]{hyperref}
\usepackage[all]{hypcap}
\usepackage{mathrsfs}

\usepackage[percent]{overpic}
\usepackage{fixltx2e}
\usepackage{epsfig}
\usepackage{verbatim}
\usepackage{multirow}
\usepackage{float}
\usepackage{array}
\usepackage{alltt}

\usepackage{threeparttable}

\newcommand{\et}{\mathrm{and}}
\newcommand{\bigo}{\mathcal{O}}

\journal{Nuclear Instruments and Methods A}
\begin{document}
\begin{frontmatter}
\title{Pulse processing routines for neutron time-of-flight data}

\author[a]{P.~\v{Z}ugec\corref{cor1}}
\ead{pzugec@phy.hr}

\author[b]{C.~Wei{\ss}}
\author[c]{C.~Guerrero}

\author[d]{F.~Gunsing}

\author[b]{V.~Vlachoudis}
\author[b,c]{M.~Sabate-Gilarte}
\author[e]{A.~Stamatopoulos}
\author[f]{T.~Wright}
\author[c]{J. Lerendegui-Marco} 
\author[g]{F.~Mingrone}
\author[f]{J.~A.~Ryan}
\author[f]{S.~G.~Warren}
\author[b,e]{A.~Tsinganis}
\author[h]{M.~Barbagallo}

\author{The n\_TOF Collaboration\fnref{fn1}}
\cortext[cor1]{Corresponding author. Tel.: +385 1 4605552}
\fntext[fn1]{www.cern.ch/ntof}

\address[a]{Department of Physics, Faculty of Science, University of Zagreb, Croatia}
\address[b]{European Organization for Nuclear Research (CERN), Geneva, Switzerland}
\address[c]{Universidad de Sevilla, Spain}
\address[d]{Commissariat \`{a} l'\'{E}nergie Atomique (CEA) Saclay - Irfu, Gif-sur-Yvette, France}

\address[e]{National Technical University of Athens (NTUA), Greece}
\address[f]{University of Manchester, Oxford Road, Manchester, UK}

\address[g]{Dipartimento di Fisica, Universit\`a di Bologna, and Sezione INFN di Bologna, Italy}
\address[h]{Istituto Nazionale di Fisica Nucleare, Bari, Italy}

%

\begin{abstract}
A pulse shape analysis framework is described, which was developed for n\_TOF-Phase3, the third phase in the operation of the n\_TOF facility at CERN. The most notable feature of this new framework is the adoption of generic pulse shape analysis routines, characterized by a minimal number of explicit assumptions about the nature of pulses. The aim of these routines is to be applicable to a wide variety of detectors, thus facilitating the introduction of the new detectors or types of detectors into the analysis framework. The operational details of the routines are suited to the specific requirements of particular detectors by adjusting the set of external input parameters. Pulse recognition, baseline calculation and the pulse shape fitting procedure are described. Special emphasis is put on their computational efficiency, since the most basic implementations of these conceptually simple methods are often computationally inefficient.
\end{abstract}


\begin{keyword}
signal analysis algorithms
\sep
pulse recognition
\sep
signal baseline
\sep
pulse shape fitting
\sep
n\_TOF facility
\end{keyword}
\end{frontmatter}

\section{Introduction}
\label{introduction}

\begin{figure*}[t!]
\centering
\includegraphics[width=0.75\linewidth,keepaspectratio]{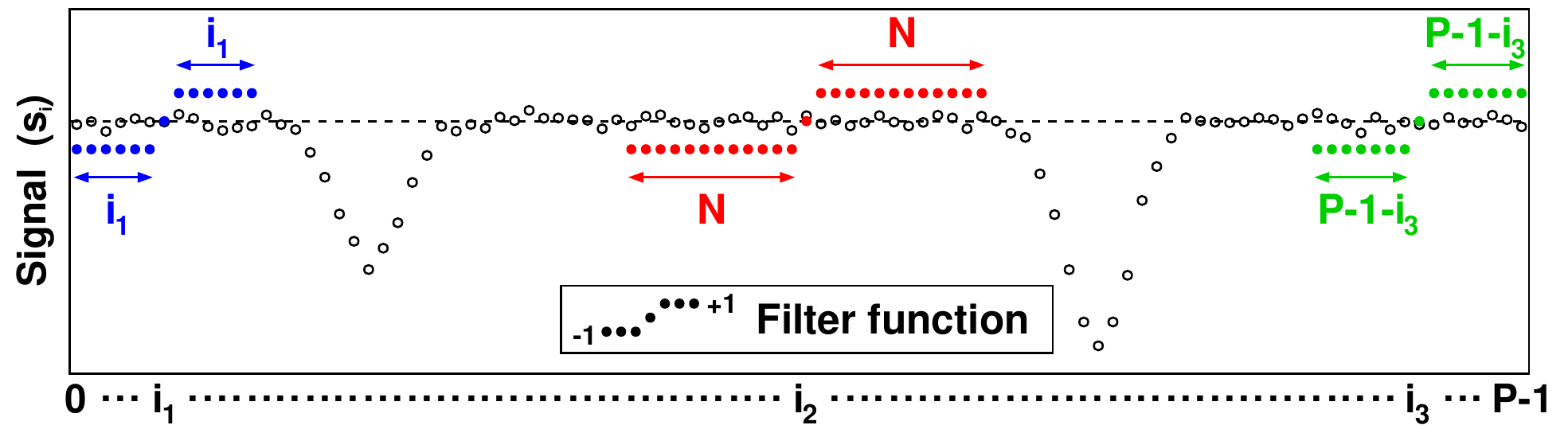}
\caption{(Color online) Illustration of the procedure for calculating the signal derivative from Eq.~(\ref{eq1}). The filter of step-size $N$ (red dots) is applied to the artificially constructed signal (open dots). The behavior of the filter at signal boundaries is shown as well (blue and green dots).}
\label{fig0}
\end{figure*}


After a year and a half long shutdown, the neutron time of flight facility n\_TOF \cite{ntof1,ntof2} at CERN has entered a third phase of its operation, known as n\_TOF-Phase3. The new era of the n\_TOF facility is marked by the successful completion of the construction of Experimental Area~2 (EAR2) \cite{ear2_0,ear2_1,ear2_2}, which was recently put into operation. Experimental Area~1 (EAR1), already in function for more than a decade, operates in parallel. The in-depth description of the general features of the n\_TOF facility, such as the neutron production and the neutron transport, may be found in Refs.~\cite{ear2_1,ear2_2,carlos}.
 
At n\_TOF a wide variety of detectors is used for measuring neutron induced reactions, including neutron capture ($n,\gamma$), neutron induced fission ($n,f$) and reactions of type ($n,p$), ($n,t$) and ($n,\alpha$). Among these are solid-state detectors  (such as the silicon based neutron beam monitor \cite{simon} and CVD diamond detectors \cite{diamond}), scintillation detectors (an array of BaF$_2$ scintillator crystals \cite{tac}, C$_6$D$_6$ liquid scintillators \cite{c6d6}) and gaseous detectors (such as MicroMegas-based detectors \cite{mgas1,mgas2}, a calibrated fission chamber from the Physikalisch Technische Bundesanstalt \cite{ptb}, a set of Parallel Plate Avalanche Counters \cite{ppac}). Several other types of detectors were recently introduced and tested at n\_TOF, such as solid-state HPGe, scintillation NaI, gaseous $^3$He detectors, etc.


A high-performance digital data acquisition system is used for the management and storage of the electronic detector signals. The system is based on flash analog-to-digital (FADC) units, recently upgraded to handle an amplitude resolution of 8 to 12 bits. It operates at sampling rates typically ranging from 100~MHz to 1~GHz, with a memory buffer of up to 175~MSamples, allowing for an uninterrupted recording of the detector output signals during the full time-of-flight range of approximately 100~ms (as used in EAR1). A detailed description of the previous version of this system can be found in Ref. \cite{daq}.

Once stored in digital form, the electronic signals have to be accessed for offline analysis, in order to obtain the time-of-flight and pulse height information for each detected pulse. The analysis procedures applied to the signals from C$_6$D$_6$ and BaF$_2$ detectors have already been described in Refs.~\cite{daq,baf2_analysis}. In order to efficiently and consistently accommodate analysis requirements of a wide variety of detectors used at n\_TOF, a generic type of routine was recently developed that can be applied to different types of signals. The routine is characterized by a minimal number of explicit assumptions about the nature of signals and is based on a pulse template adjustment, which we refer to as the \emph{pulse shape fitting}. For each detector or type of detector a set of analysis parameters needs to set externally. A number of these will be mentioned throughout this paper.

Many of the procedures adopted for the signal analysis -- such as the pulse integration with the goal of extracting the energy deposited in the detectors, or the constant fraction discrimination for determining the pulses' timing properties -- are all well established techniques, thus we don't find it necessary to enter their description. Consequently, we will focus on the technical aspects of the more specific methods that were found to perform very well for the wide variety of signals from different detectors, in order to provide their documentation and ensure their reproducibility. Special emphasis will be put on the computational efficiency of these procedures.

Selected examples of the signals from the detectors available at n\_TOF are shown throughout the paper. Regarding the previous works on the signal analysis procedures adapted to the specific types of detectors, the reader may consult Refs.~\cite{psa_naid,psa_hpge,psa_sili,psa_scint}, dealing with NaI, HPGe, silicon and organic scintillation detectors, respectively. We also refer the reader to an exhaustive comparative analysis of many different pulse shape processing methods comprehensively covered in Ref.~\cite{psa_review}, and to the references contained therein.


\section{Pulse recognition}
\label{recognition}

\subsection{Signal derivative}

The central procedure in the pulse recognition is the construction of the signal derivative $d$. We use the following definition:
\begin{linenomath}\begin{equation}
\label{eq1}
d_i\equiv\sum_{j=1}^{\min[N,i,P-1-i]}\hspace*{-5mm}(s_{i+j}-s_{i-j})
\end{equation}\end{linenomath}
that takes advantage of integrating the signal $s$ at both sides of the $i$-th point at which the derivative is to be calculated. $P$ is the total number of points composing the recorded signal. The points are enumerated from 0 to $P-1$, which is a convention used throughout this paper, unless explicitly stated otherwise. A \emph{step-size} $N$ is the default number of points to be taken for integration. As illustrated by Fig.~\ref{fig0}, this procedure formally resembles a convolution between the signal and a see-saw-shaped filter function of unit height, up to the boundary effects regulated by the upper summation bound from Eq.~(\ref{eq1}). Evidently, when the step-size $N$ is adjusted so as to be wider than the period of noise in the signal (and narrower than the characteristic pulse length), the procedure acts as a low-pass filter, improving the signal-to-noise ratio in the derivative.

The number of operations required by the straightforward implementation of this algorithm is proportional to $N\times P$, making such approach computationally inefficient. Fortunately, recursive relations may be derived for calculating the consecutive $d_i$ terms, making the entire procedure linear in the number of required operations: $\bigo(P)$. By defining the forward and backward sums $\Sigma_i^+$ and $\Sigma_i^-$, respectively, as:
\begin{linenomath}\begin{equation}
\label{eq2}
\Sigma_i^\pm\equiv\sum_{j=1}^{\min[N,i,P-1-i]}\hspace*{-5mm}s_{i\pm j}
\end{equation}\end{linenomath}
the derivative may be rewritten as: $d_i=\Sigma_i^+-\Sigma_i^-$. The initial values $\Sigma_0^+=0$ and $\Sigma_0^-=0$ follow directly from
Eq.~(\ref{eq1}). The recursive relations for subsequent pairs of $\Sigma_i^+$ and $\Sigma_i^-$ are given in Table~\ref{tab1}, being listed according to the boundary effects.

\begin{table}[b!]
\caption{List of recursive relations for calculating forward and backward sums $\Sigma_i^+$ and $\Sigma_i^-$ from Eq. (\ref{eq2}). The signal derivative $d_i$ may then be obtained as: $d_i=\Sigma_i^+-\Sigma_i^-$. Cases are categorized based on the boundary effects (whether the integration windows defined by the step-size $N$ reach the boundaries of the waveform, composed of total of $P$ points), as illustrated in Fig.~\ref{fig0}.}
\label{tab1}
\centering
\begin{tabular}{l}
\hline\hline
\textbf{Beginning of the waveform}\\
$\boldsymbol{i\le N\qquad\et\qquad 2i\le P-1}$\\
\hline
$\Sigma_i^+=\Sigma_{i-1}^+ -s_i+s_{2i-1}+s_{2i}$\\
$\Sigma_i^-=\Sigma_{i-1}^- +s_{i-1}$\\
\hline
\textbf{Middle of the waveform}\\
$\boldsymbol{i> N\qquad\et\qquad i+N\le P-1}$\\
\hline
$\Sigma_i^+=\Sigma_{i-1}^+ -s_i+s_{i+N}$\\
$\Sigma_i^-=\Sigma_{i-1}^- +s_{i-1}-s_{i-1-N}$\\
\hline
\textbf{End of the waveform}\\
$\boldsymbol{i+N> P-1\qquad\et\qquad 2i> P-1}$\\
\hline
$\Sigma_i^+=\Sigma_{i-1}^+ -s_i$\\
$\Sigma_i^-=\Sigma_{i-1}^- +s_{i-1}-s_{2i-P-1}-s_{2i-P}$\\
\hline\hline
\end{tabular}
\end{table}

\subsection{Derivative crossing thresholds}

\begin{figure}[t!]
\includegraphics[width=1.\linewidth,keepaspectratio]{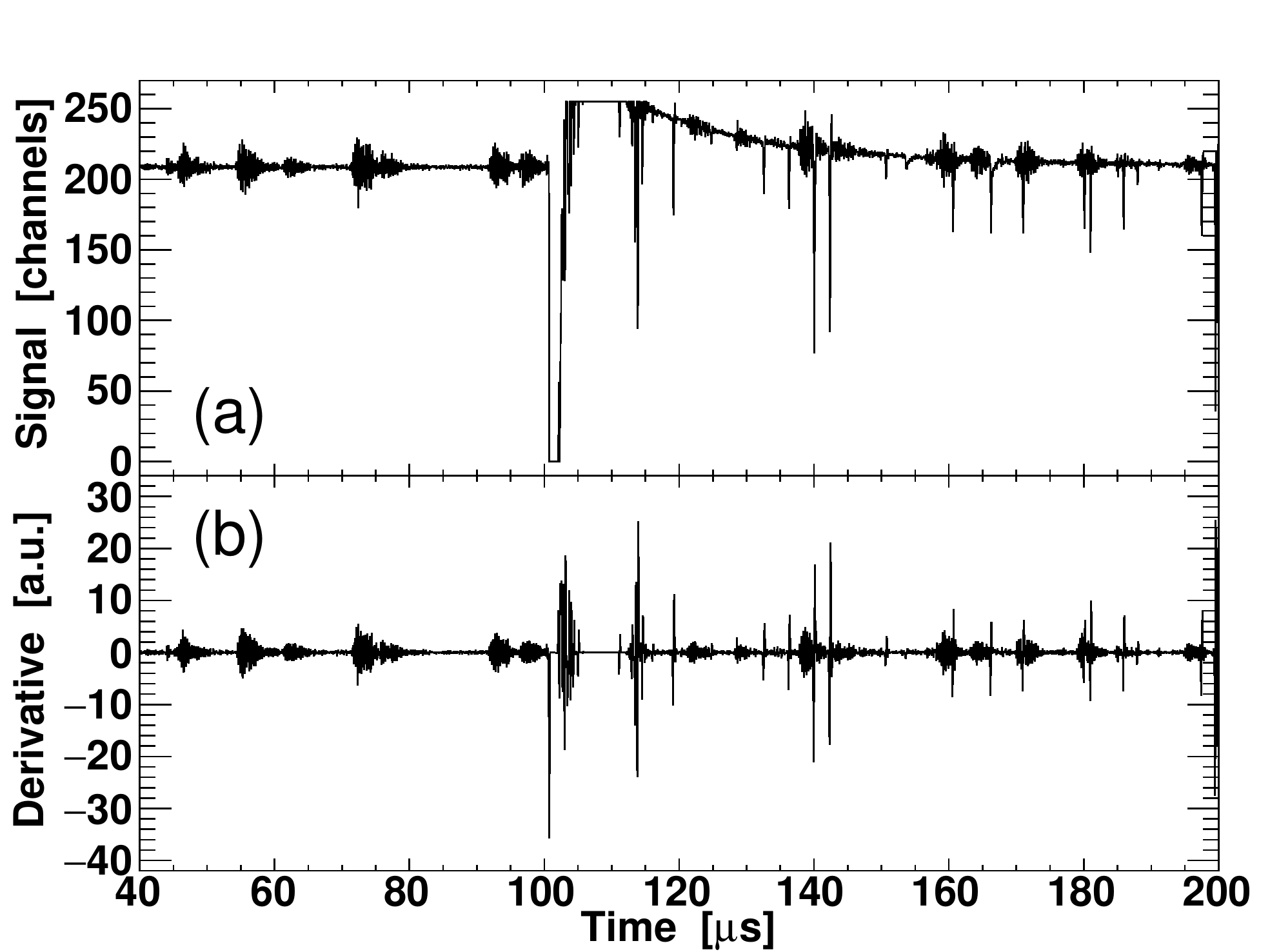}
\caption{Top panel (a): example of the digitized signal from MicroMegas detector. Bottom panel (b): its derivative calculated from Eq. (\ref{eq1}).}
\label{fig1}
\end{figure}

In order to recognize the presence of the pulses in the overall signal, their derivative must cross certain predefined thresholds. These thresholds need to be set in such a way as to reject most of the noise, but not to discard even the lowest pulses. Therefore, they must be adaptively brought into connection with the level of the noise characteristic of the current waveform, which is best expressed through the root mean square (RMS) of the noise. Figure~\ref{fig1} shows an example of one of the most challenging signals for this task, the signal from a MicroMegas detector. Top panel (a) shows the selected fraction of an actual recorded signal, with the strongest pulse corresponding to an intense $\gamma$-flash caused by the proton beam hitting the spallation target, while the bottom (b) panel shows its derivative calculated from Eq.~(\ref{eq1}). This signal is heavily affected by the random beats which do not qualify as the pulses of interest to any meaningful measurement (by \emph{beats} we consider the coherent noise resembling acoustic beats, as shown in Fig.~\ref{fig1} and later in Fig.~\ref{fig8}). Several tasks are immediately evident. First, the pulses themselves must be excluded from the procedure for determining the derivative thresholds, since they can only increase the overall RMS, thus leading to a rejection of the lowest pulses. However, the pulses can not be discriminated from the noise before the thresholds have been found. Second, the beats must not be assigned to the noise RMS, since they are only sporadic and can also only lead to an unwanted increase in thresholds. Finally, in some cases one can not even rely on the assumption of a fixed number of clear presamples before the first significant pulse, such as the initial $\gamma$-flash pulse. This is the case in measurements with high activity samples, when their natural radioactivity causes a continual stream of pulses, independent of the external experimental conditions. Another example is the intake of waveforms for certain calibration purposes, when no external trigger is used and signals are recorded without any guarantee of clear presamples. In order to meet all these challenges, the procedure of applying the weighted fitting to the modified distribution of derivative points. It may be decomposed into four basic steps, described throughout this section.

\textbf{Step 1:} build the distribution (histogram) of all derivative points. As Fig. \ref{fig1} shows, all the points from the derivative baseline are expected to group around the value 0, forming a peak characterized by the RMS of the noise. On the other hand, the points from the sporadic pulses and/or beats are expected to form the long tails of the distribution. Since the central peak of the distribution carries the information about the sought for RMS, it needs to be reconstructed by means of (weighted) fitting. 

A technicality is related to the treatment of the central bin, corresponding to the derivative value 0. It has been observed that in certain cases an excessive number of points is accumulated in this bin, making it reach out high above the rest of the distribution. Depending on the specific signal conditions, this feature has proven to be either beneficial or detrimental to the quality of the fitting procedure. Therefore, the content $N_c$ of the central ($c$-th) bin is replaced by:
\begin{linenomath}\begin{equation}
N_c\quad\rightarrow\quad \sqrt{N_c\times\frac{N_{c-1}+N_{c+1}}{2}}
\end{equation}\end{linenomath}
i.e. by the geometrical mean between the initial content and the arithmetic mean of the neighboring bins. Since the geometric mean is biased towards the smaller of the averaged terms, this solution was selected in an attempt of finding an ideal compromise between retaining the signature of the original bin content and bringing it down towards the main fraction of the histogram. It was found that after this modification the RMS of the fitted distribution is very well adjusted to the derivative baseline in both cases: when the initial bin content would have worked either to the advantage or the detriment of the fitting procedure.

\textbf{Step 2:} adjust the histogram range. After building the initial distribution, taking into account all derivative points and adjusting the central bin, the histogram range is reduced by cutting it symmetrically around 0 until 10\% of its content has been discarded. This procedure helps in localizing the relevant part of the distribution by rejecting the sporadic far-away points, thus limiting the range of the distribution from $-d_\mathrm{max}$ to $d_\mathrm{max}$, which will be of central importance in defining the weights for the weighted fitting.

\textbf{Step 3:} emphasize the central peak. One must consider that even with appropriate weights, the fitting might still be heavily affected by the long tails of the distribution, increasing the final extracted RMS. In order to compensate for this effect, the central peak is better pronounced by exponentiating the entire distribution, i.e. by replacing the content $N_i$ of the $i$-th histogram bin by the following value:
\begin{linenomath}\begin{equation}
N_i\quad\rightarrow\quad e^{N_i}-1
\end{equation}\end{linenomath}
This procedure affects the width of the central peak, narrowing it somewhat when there are no significant tails. The lower extracted RMS is preferred over the higher one, in order for the derivative thresholds not to reject the lowest pulses. As it will be explained later, the accidental triggering of lower thresholds by the noise will be discarded by the appropriate pulse elimination procedure. Before exponentiating the histogram content, care must be taken to rescale it appropriately -- e.g. by scaling the distribution peak to unity -- in order to avoid the potential numerical overflow. Furthermore, a consistent normalization is crucial in making the procedure insensitive to the length of the recorded signal (i.e. the initial height of distribution), since the exponentiation is nonlinear in the absolute number of counts $N_i$.

\begin{figure}[t!]
\includegraphics[width=1.\linewidth,keepaspectratio]{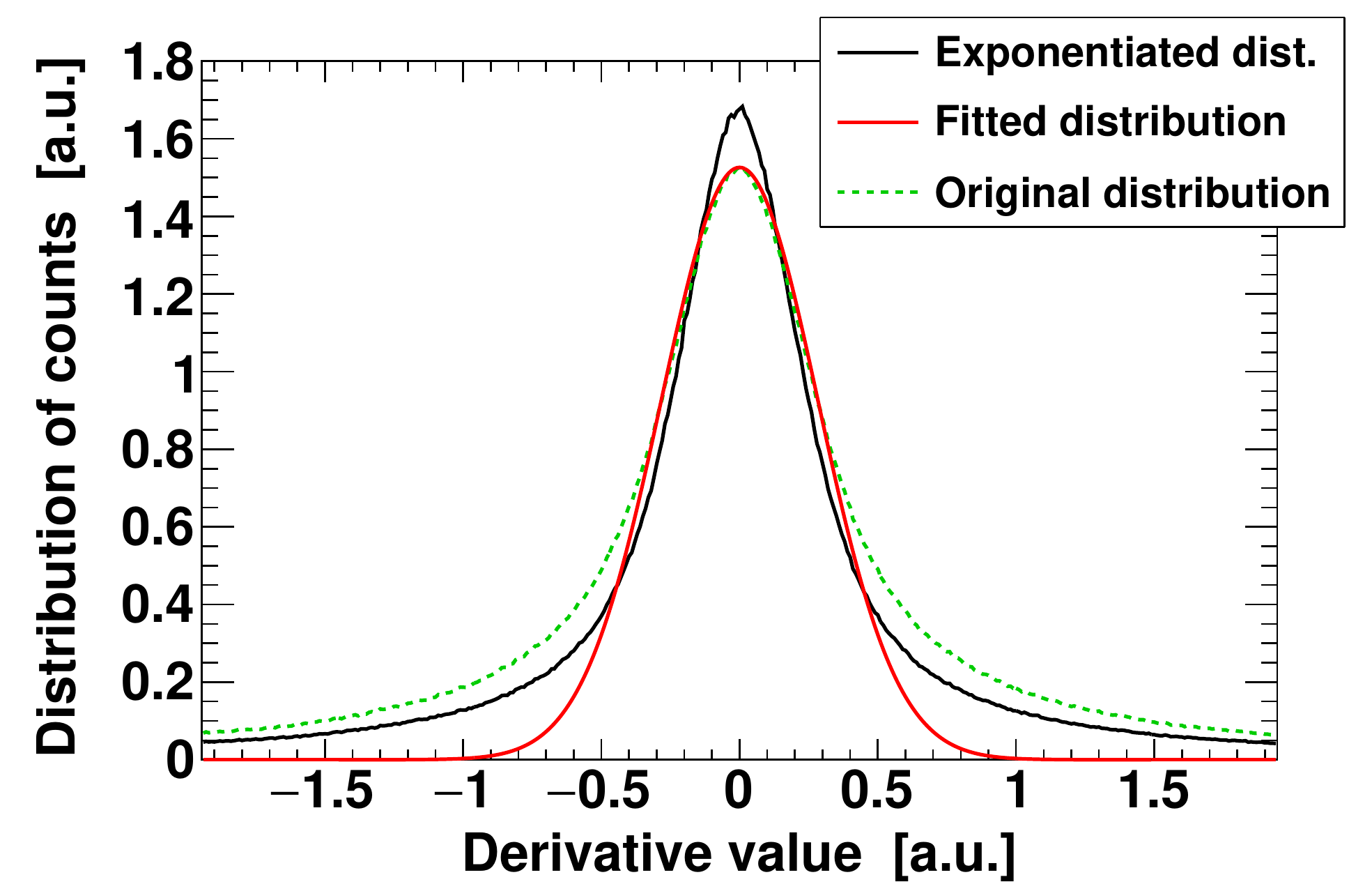}
\caption{(Color online) Distribution of derivative points from Fig.~\ref{fig1}, with the result of the weighted fitting designed to reconstruct the width of the central peak. The dashed line shows the true distribution of points from Fig.~\ref{fig1}, arbitrarily scaled to a height of a fitted distribution. The exponentiated distribution was obtained starting from the original distribution scaled to unity.}
\label{fig2}
\end{figure}

\textbf{Step 4:} perform the weighted fitting so as to best reconstruct the central peak. The remaining distribution is fitted to a Gaussian shape explicitly assumed to be centered at 0, by minimizing the following expression:
\begin{linenomath}\begin{equation}
\label{eq5}
\sum_{i=i_\mathrm{min}}^{i_\mathrm{max}}W_i\left(N_i-Ae^{-x_i^2/(2\Delta^2)}\right)^2
\end{equation}\end{linenomath}
where $x_i$ is the abscissa coordinate of the $i$-th bin, such that $x_{i_\mathrm{min}}=-d_\mathrm{max}$ and $x_{i_\mathrm{max}}=d_\mathrm{max}$. Parameters $A$ and $\Delta$ are to be determined by fitting. At the end of the procedure, $\Delta$ is identified with the RMS of the central peak, i.e. with the RMS of the noise in the derivative. The selection of a Gaussian as a prior is justified by the Central Limit Theorem, applied to a sum of random noise values from Eq.~(\ref{eq1}). Central to the fitting are the weights $W_i$, which have been selected to follow the Gaussian dependence:
\begin{linenomath}\begin{equation}
W_i=e^{-x_i^2/(2\Lambda^2)}
\end{equation}\end{linenomath}
with a standard deviation $\Lambda$. By an empirical optimization it was set to \mbox{$\Lambda=d_\mathrm{max}/4$}. These weights efficiently suppress the impact from the tails of the distribution, while giving precedence to the central peak. For the fitting a Levenberg-Marquardt algorithm was adopted, as described in Ref.~\cite{numc}. Figure \ref{fig2} shows the distribution of derivative points from Fig.~\ref{fig1}, together with the central peak reconstruction by means of the weighted fitting.

While the weighted fitting is beneficial for rejecting the long tails of the distribution, the unweighted fitting has been found more appropriate for very narrow distributions covering only a few histogram bins. Due to the low number of bins and rapidly decreasing weighting factors, the weighted fitting procedure is then sensitive only to the narrow top of the distribution, which is effectively treated as flat, yielding an outstretched fit. Therefore, the unweighted fitting to the Gaussian shape from Eq.~(\ref{eq5}) is also performed. In addition, the RMS of the distribution is calculated directly  as: \mbox{$\mathrm{RMS}^2=\sum_{i=i_\mathrm{min}}^{i_\mathrm{max}}N_i x_i^2/\sum_{i=i_\mathrm{min}}^{i_\mathrm{max}}N_i$}. The lowest of the three results -- from the weighted fitting, unweighted fitting and the direct calculation -- is kept as the final one. The additional fitting and the direct calculation also serve as a contingency in case either of the fitting procedures fails to properly converge.

\subsection{Pulse discrimination}

From the derivative noise RMS extracted by one of the previously described procedures, the default values for the derivative crossing thresholds have been selected as $\pm3.5\times$RMS, due to the fact that this range corresponds to 99.95\% confidence interval under the assumption of normally distributed noise. Since the order of crossing these thresholds (together with some later analysis procedures) depends on the pulse polarity, all signals are treated as negative. This means that the signals are inverted, i.e. multiplied by $-1$, if expected to be positive from an external input parameter.

Differentiating the unipolar pulse leads to a bipolar pulse in the derivative. Therefore, the derivative of the negative unipolar pulse must, ideally, make 4 threshold crossings in this exact order: lower-lower-upper-upper. However, in case of the lowest pulses or very high pileup, the integration procedure from Eq.~(\ref{eq1}) may flatten the final derivative, not causing the second threshold crossing. Hence, the principle of 4 threshold crossings was relaxed in order to facilitate the recognition of these pulses. Thus, crossing a single threshold suffices to trigger the pulse recognition. However, if both thresholds are crossed in the order lower-upper, a single pulse is recognized, instead of two. In summary, these are the threshold crossing possibilities that mark the presence of the pulse: lower-lower (without the subsequent upper crossing), upper-upper (without the previous lower crossing) and lower-lower-upper-upper. After initially locating the pulses between the points of the first and the last threshold crossing, their range is further extended until the derivative reaches 0 at both sides, unless there are neighboring pulses in line preventing the expansion.

\begin{figure}[t!]
\includegraphics[width=1.\linewidth,keepaspectratio]{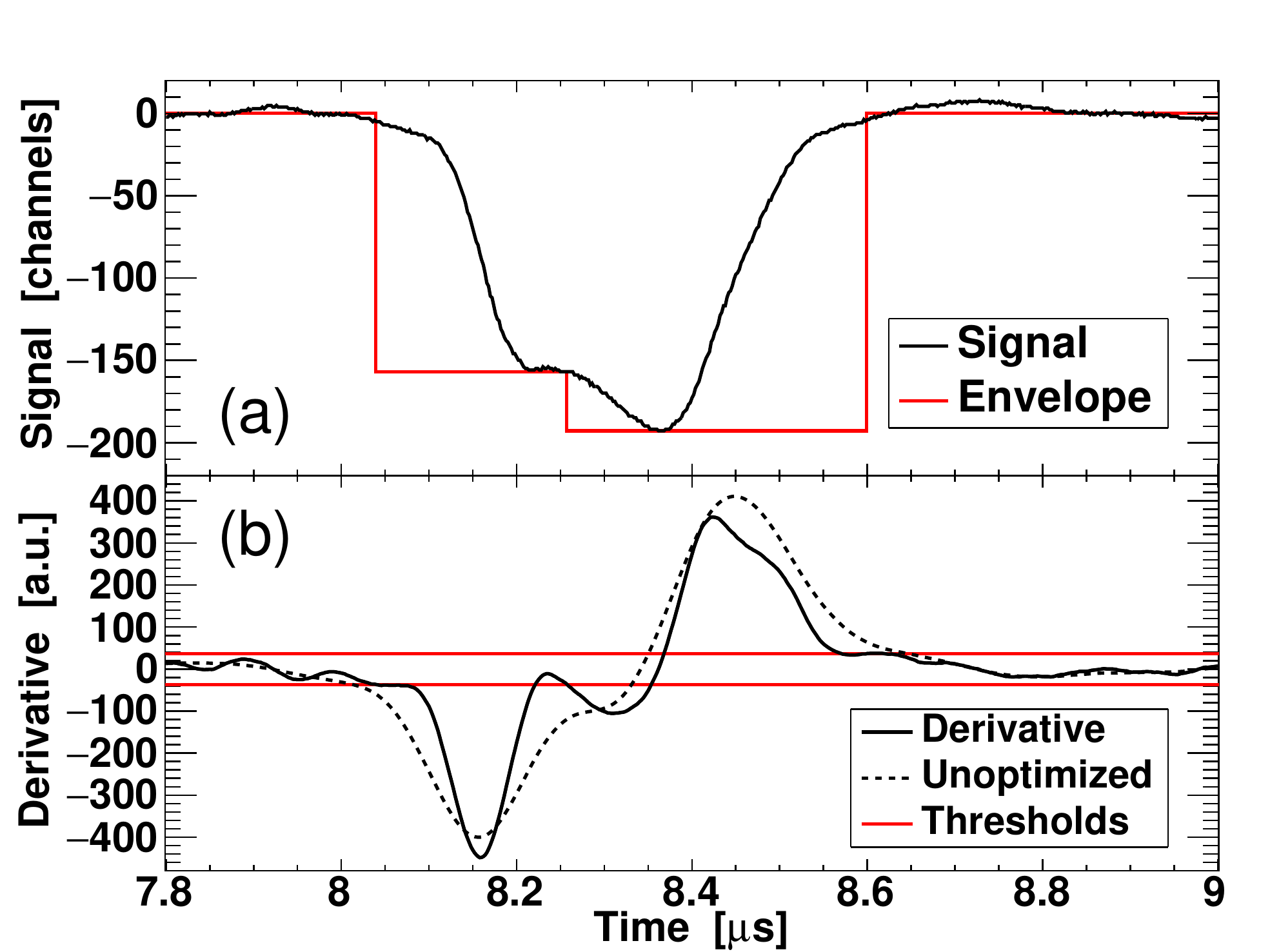}
\caption{(Color online) Pulse recognition procedure applied to the piled-up pulses. Top panel (a) shows the actual signal, with the red envelope marking the successful separation of the pulses. Bottom panel (b) shows the signal derivative crossing the appropriate thresholds and triggering the pulse recognition from panel (a). Derivative calculated with an unoptimized (too large) step-size is also shown.}
\label{fig3}
\end{figure}

The thresholds being low enough not to reject the lowest pulses will, from time to time, be accidentally triggered by the noise. These occurrences are dealt with by a set of elimination conditions, which are determined by means of the external input parameters. These conditions include the lower and upper limit for the pulse width, the lower limit for the pulse amplitude and the lower and upper limit for the area-to-amplitude ratio. The first elimination, based only on the pulse width, is performed immediately after the pulse recognition procedure. The final elimination, based on the pulse amplitudes and areas, may only be performed at a later stage, after the signal baseline has been calculated. However, it is paramount that the first stage of elimination be performed at this point, since several later procedures, such as the baseline calculation, depend on the reported pulse candidates. In case of an excessive number of falsely recognized pulses, the quality of procedures relying on the reported pulse positions may be compromised.

Figure~\ref{fig3} shows an example of a demanding case of pileup, where two pulses are successfully resolved. The top panel (a) shows the actual signal, with the red envelope confining separate pulses. The bottom panel (b) shows the optimized signal derivative crossing the thresholds, triggering the pulse recognition. It also illustrates the importance of optimizing the step-size for calculating the derivative from Eq.~(\ref{eq1}), since a further increase in step-size (dashed line) would flatten the derivative at the point of the second crossing, preventing the separation of two pulses from panel (a). For visual purposes, two displayed derivatives were normalized so that their thresholds coincide.

The described pulse recognition technique was found to perform very well for signals from a wide variety of detectors in use at n\_TOF. The example from Fig.~\ref{fig3} confirms that with optimized parameters the procedure is able to resolve quite demanding pileups. Due to the relaxed threshold crossing conditions, it is also quite sensitive even to the lowest pulses, barely exceeding the level of the noise. Since the same sensitivity characterizing the pulse recognition procedure sporadically leads to an accidental threshold crossing due to noise, an elimination procedure has been implemented alongside it.

\subsection{Multiple polarities}

The adopted pulse recognition procedure lends itself easily to signals that exhibit pulses of both polarities. In this case two derivative passes should be made -- one over the regular derivative, one over the inverted one (multiplied by $-1$). Quite often, the reported pulse candidates from two passes will overlap, since the part of a real pulse from one pass will act as a false candidate within the other pass. The pulse candidates from two passes should be analyzed independently and then submitted to the pulse elimination algorithm. It was observed that even the quite relaxed elimination conditions successfully reject the false candidates from the selection of overlapping pulses.

\subsection{Bipolar pulses}

An additional pulse range adjustment procedure was implemented in order to accommodate bipolar pulses. Since the end of the pulse is determined by the derivative reaching 0 after the first unipolar part of the pulse, the recognition of bipolar pulses stops at the extremum of the second pole. However, once the signal baseline has been calculated, the boundary of the pulse may be shifted towards the point of the baseline crossing, keeping only the first pole of the pulse or fully covering both of them. In case of two immediate but not piled-up bipolar pulses, the first one ends at the extremum of its own second pole, where the next pulse is immediately recognized to start, due to the behavior of the derivative $d$. Therefore, the starting points of the pulses need to be adjusted (with respect to the calculated baseline) in accordance with the requirements of a specific signal, so that the finally determined range of the second pulse does not start prematurely, preventing also the (optional) expansion of the first pulse.

\section{Baseline}
\label{baseline}

Three different baseline methods have been implemented, that may all be used within the same waveform, depending on the signal behavior. These are the constant baseline, the weighted moving average and the moving maximum. The use of the moving maximum is usually related only to the first part of the waveform, when the effect of the $\gamma$-flash upon the signal is extreme (there is also an alternative method of subtracting the baseline distortion pulse shape, designed for this region). Moving average is also related to the baseline distortion by $\gamma$-flash, however it is often the most appropriate method to be used throughout the entire waveform, especially if the baseline exhibits slow oscillations. Constant baseline is suitable only after the baseline has been fully restored after the initial $\gamma$-flash, or if the detector response to external influences is remarkably stable.

\subsection{Constant baseline}

A constant baseline is calculated as the average of all signal points between the pulse candidates reported by the pulse recognition procedure. In this way any need for an iterative procedure is avoided, while the baseline remains unaffected by the actual pulses.

\subsection{Weighted moving average}

\begin{table}[t!]
\caption{List of recursive relations for evaluating the baseline from Eq.~(\ref{eq3}). The involved terms are defined by Eq.~(\ref{eq11}). The separate cases refer to the position of the averaging window, defined by the window parameter $N$, relative to the edges of the waveform composed of $P$ points. Constants $\kappa_\mathrm{c}\equiv\cos\tfrac{\pi}{N}$ and $\kappa_\mathrm{s}\equiv\sin\tfrac{\pi}{N}$ have been introduced for the efficiency of the calculation.}
\label{tab2}
\centering
\begin{tabular}{l}
\hline\hline
\textbf{Window protrudes at the beginning of the waveform}\\
$\boldsymbol{i-N\le0\qquad\et\qquad i+N\le P-1}$\\
\hline
$K_i^{(\lambda)}=K_{i-1}^{(\lambda)}+\lambda_{i+N}$\\
$C_i^{(\lambda)}=\kappa_\mathrm{c}\times C_{i-1}^{(\lambda)}+\kappa_\mathrm{s}\times S_{i-1}^{(\lambda)}-\lambda_{i+N}$\\
$S_i^{(\lambda)}=\kappa_\mathrm{c}\times S_{i-1}^{(\lambda)}-\kappa_\mathrm{s}\times C_{i-1}^{(\lambda)}$\\
\hline
\textbf{Window is contained within the waveform}\\
$\boldsymbol{i-N>0\qquad\et\qquad i+N\le P-1}$\\
\hline
$K_i^{(\lambda)}=K_{i-1}^{(\lambda)}-\lambda_{i-1-N}+\lambda_{i+N}$\\
$C_i^{(\lambda)}=\kappa_\mathrm{c}\times C_{i-1}^{(\lambda)}+\kappa_\mathrm{s}\times S_{i-1}^{(\lambda)}+\kappa_\mathrm{c}\times \lambda_{i-1-N}-\lambda_{i+N}$\\
$S_i^{(\lambda)}=\kappa_\mathrm{c}\times S_{i-1}^{(\lambda)}-\kappa_\mathrm{s}\times C_{i-1}^{(\lambda)}-\kappa_\mathrm{s}\times \lambda_{i-1-N}$\\
\hline
\textbf{Window protrudes at the end of the waveform}\\
$\boldsymbol{i-N>0\qquad\et\qquad i+N> P-1}$\\
\hline
$K_i^{(\lambda)}=K_{i-1}^{(\lambda)}-\lambda_{i-1-N}$\\
$C_i^{(\lambda)}=\kappa_\mathrm{c}\times C_{i-1}^{(\lambda)}+\kappa_\mathrm{s}\times S_{i-1}^{(\lambda)}+\kappa_\mathrm{c}\times \lambda_{i-1-N}$\\
$S_i^{(\lambda)}=\kappa_\mathrm{c}\times S_{i-1}^{(\lambda)}-\kappa_\mathrm{s}\times C_{i-1}^{(\lambda)}-\kappa_\mathrm{s}\times \lambda_{i-1-N}$\\
\hline
\textbf{Window protrudes at both ends of the waveform}\\
$\boldsymbol{i-N\le0\qquad\et\qquad i+N> P-1}$\\
\hline
$K_i^{(\lambda)}=K_{i-1}^{(\lambda)}$\\
$C_i^{(\lambda)}=\kappa_\mathrm{c}\times C_{i-1}^{(\lambda)}+\kappa_\mathrm{s}\times S_{i-1}^{(\lambda)}$\\
$S_i^{(\lambda)}=\kappa_\mathrm{c}\times S_{i-1}^{(\lambda)}-\kappa_\mathrm{s}\times C_{i-1}^{(\lambda)}$\\
\hline\hline
\end{tabular}
\end{table}

The moving average is the appropriate method for determining the baseline whenever the clear information about the baseline is, in fact, available, i.e. when the uninterrupted portions of the baseline may indeed be found within the signal. The following definition is used for the weighted moving average:
\begin{linenomath}\begin{equation}
\label{eq3}
B_i\equiv\frac{\displaystyle\sum_{j=\max[0,i-N]}^{\min[i+N,P-1]}\hspace*{-5mm}s_jw_j\big\{1+\cos\big[(j-i)\tfrac{\pi}{N}\big]\big\}}{\displaystyle\sum_{j=\max[0,i-N]}^{\min[i+N,P-1]}\hspace*{-5mm}w_j\big\{1+\cos\big[(j-i)\tfrac{\pi}{N}\big]\big\}}
\end{equation}\end{linenomath}
with $N$ as the number of points (referred to as the \emph{window parameter}) at each side of the $i$-th one to be taken for averaging the signal $s$, composed of the total of $P$ points. It should be noted that the averaging window is $2N+1$ points wide. The weighting kernel is given by the cosine (i.e. Hann \cite{hann}) window, with additional weighting factors $w_i$ that are equal to the number of uninterrupted points within a given stretch of the baseline. Inside the reported pulse candidates, these weights should be much lower than unity ($w_i\ll1$), so as to exclude the pulses from the baseline calculation. However, a finite non-zero value is required, in order to avoid division by zero in case the averaging window is completely contained within the pulse. For the weighting factors inside the pulses we have adopted the value $10^{-6}$. More precisely, for $Q$ as the total number of pulses identified inside the waveform -- with $\alpha_q$ and $\beta_q$ denoting the first and the last index of the $q$-th pulse ($q\in[1,Q]$), respectively -- the weighting factors are defined as:
\begin{linenomath}\begin{equation}
w_i=\left\{\begin{array}{cl}
10^{-6}&\mathrm{if}\;\;i\in[\alpha_q,\beta_q]\;\;\mathrm{for\;any}\;\;q\in[1,Q]\\
\alpha_{q+1}-\beta_q-1&\mathrm{if}\;\;i\in\langle\beta_q,\alpha_{q+1}\rangle\;\;\mathrm{for\;any}\;\;q\in[0,Q]
\end{array}\right.
\end{equation}\end{linenomath}
where $\beta_0=-1$ and $\alpha_{Q+1}=P$. Evidently, the window parameter $N$, given as an external parameter, should be large enough to connect the baseline at both sides of the widest pulse or the widest expected chain of piled-up pulses. The initial elimination of falsely recognized pulses (based on their widths) also plays a role in this procedure, since every reported pulse interrupts the baseline, affecting the weighting factors $w$. Still, the procedure is quite robust against this change of the weighting factors.

\begin{figure}[b!]
\includegraphics[width=1.\linewidth,keepaspectratio]{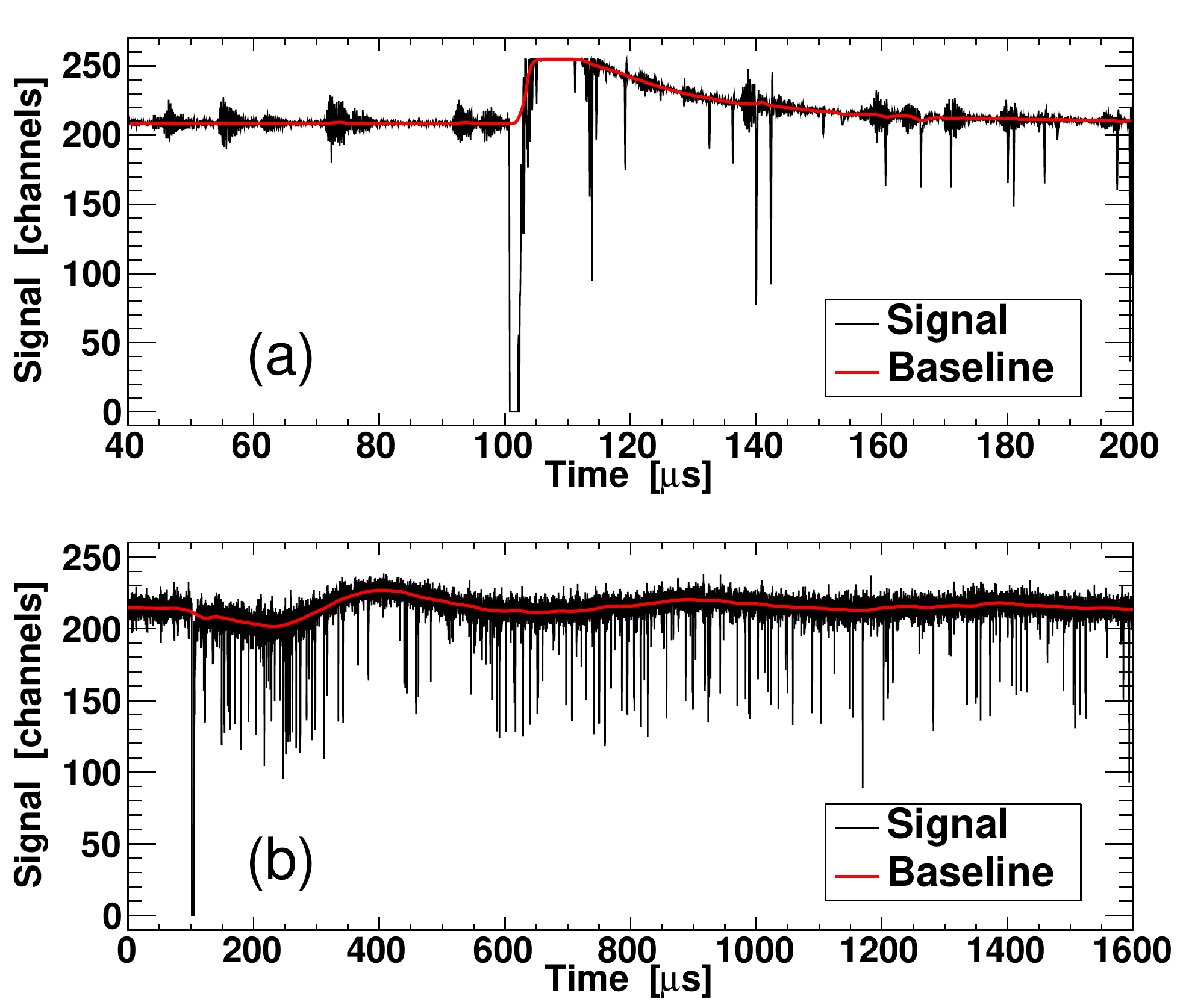}
\caption{(Color online) Independent examples of the adaptive baseline calculated using the weighted moving average procedure from Eq.~(\ref{eq3}).}
\label{fig4}
\end{figure}

The form of the summation bounds from Eq.~(\ref{eq3}) properly takes into account the boundary cases, when the averaging window reaches the edges of the signal. Once again, the straightforward implementation of the algorithm for evaluating Eq.~(\ref{eq3}) is of $\bigo(N\times P)$ computational complexity. Hence,  recursive relations have been derived, which provide a linear dependence in the number of operations for calculating the baseline throughout the entire waveform: $\bigo(P)$. We define the following terms:
\begin{linenomath}\begin{align}
\label{eq11}
\begin{split}
K_i^{(\lambda)}&\equiv\sum_{j=\max[0,i-N]}^{\min[i+N,P-1]}\hspace*{-5mm}\lambda_j\\
C_i^{(\lambda)}&\equiv\sum_{j=\max[0,i-N]}^{\min[i+N,P-1]}\hspace*{-5mm}\lambda_j\cos\big[(j-i)\tfrac{\pi}{N}\big]\\
S_i^{(\lambda)}&\equiv\sum_{j=\max[0,i-N]}^{\min[i+N,P-1]}\hspace*{-5mm}\lambda_j\sin\big[(j-i)\tfrac{\pi}{N}\big]
\end{split}
\end{align}\end{linenomath}
allowing to rewrite Eq.~(\ref{eq3}) as:
\begin{linenomath}\begin{equation}
B_i=\frac{K_i^{(sw)}+C_i^{(sw)}}{K_i^{(w)}+C_i^{(w)}}
\end{equation}\end{linenomath}
where the notation $\lambda=sw$ implies: $\lambda_j=s_jw_j$. Initial values $K_0^{(\lambda)}$, $C_0^{(\lambda)}$ and $S_0^{(\lambda)}$ are to be calculated directly from Eq.~(\ref{eq3}). The recursive relations for calculating all subsequent terms are listed in Table~\ref{tab2}, according to the position of the averaging window, relative to the waveform boundaries. It should be noted that the efficient calculation requires the terms $\cos\tfrac{\pi}{N}$ and $\sin\tfrac{\pi}{N}$ to be treated as constants and calculated only once, instead of repeating the calculation at each step. Figure \ref{fig4} shows two examples of the performance of the described baseline procedure.

\subsection{Moving maximum}
\label{envelope}

\begin{figure}[t!]
\includegraphics[width=1.\linewidth,keepaspectratio]{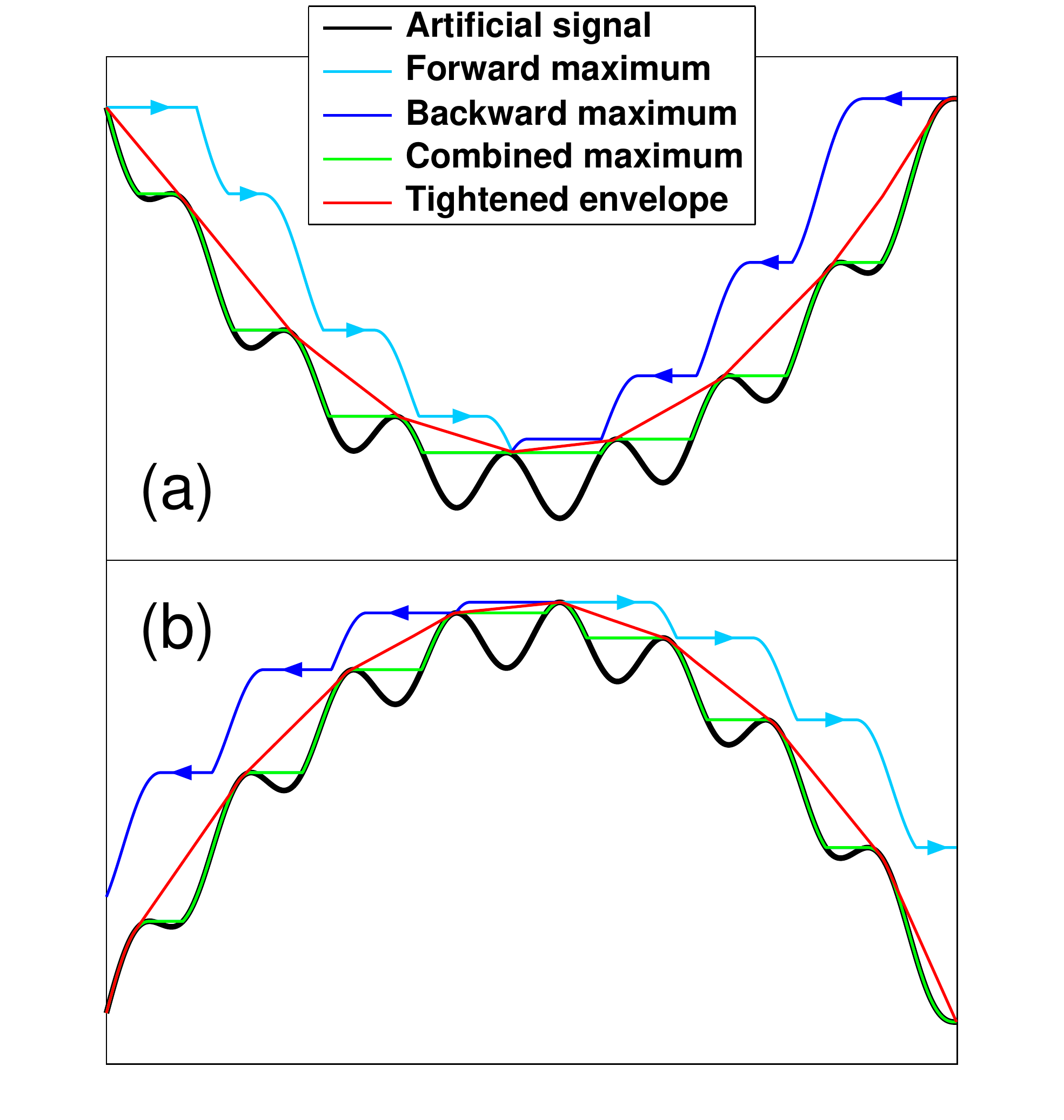}
\caption{(Color online) Proof of concept for finding the upper signal envelope by combining the forward and backward moving maximum. The tightened envelope is also shown. The signals have been artificially constructed.}
\label{fig5}
\end{figure}

\begin{figure}[t!]
\includegraphics[width=1.\linewidth,keepaspectratio]{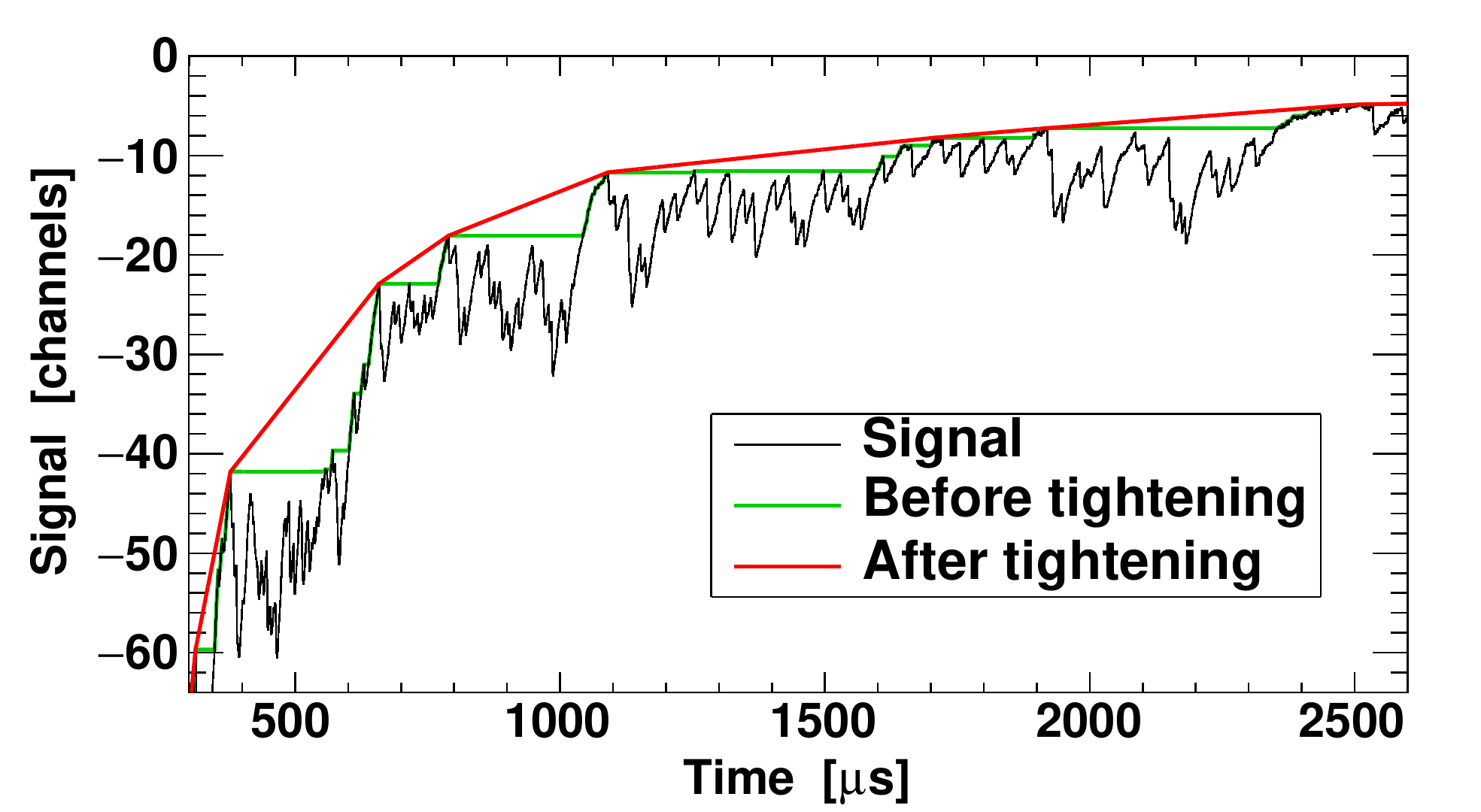}
\caption{(Color online) Example of the signal from a gaseous $^{3}$He detector, that requires the reconstruction of the upper envelope in order to identify the baseline. The envelope is shown both before and after the tightening procedure.}
\label{fig6}
\end{figure}

The following baseline procedure is appropriate when the information about the signal baseline has been (almost) completely lost due to the sequential and persistent pileup of pulses, while the baseline itself is known not to be constant and no other \emph{a priori} knowledge about it is available  (example given later in Fig.~\ref{fig6}). In this case the best, if not the only assumption to be made is that the baseline follows the signal envelope, defined by the dips between the pulses, especially those that manage to reach most deeply toward the true baseline. Since all signals are treated as negative, as stated before, the upper envelope needs to be found. This may me done by constructing two moving maxima -- one that we refer to as the \emph{forward maximum}, the other as the \emph{backward maximum} -- and taking the minimum of the two at each point of the signal (the advantages of this kind of competitive approach have already been explored in the past \cite{competitive}). We define the forward maximum at \mbox{$i$-th} point as the maximal signal value from a \emph{moving window} of $N$ points before the $i$-th one, with backward maximum as the maximal value from the window of $N$ points after the $i$-th one:
\begin{linenomath}\begin{align}
\begin{split}
&M_i^{(\mathrm{forward})}=\mathrm{max}[s_{\mathrm{max}[0,i-N+1]},\ldots,s_i]\\
&M_i^{(\mathrm{backward})}=\mathrm{max}[s_i,\ldots,s_{\mathrm{min}[i+N-1,P-1]}]
\end{split}
\end{align}\end{linenomath}
As before, $P$ is the total number of points in the waveform, with $N$ as the external input parameter. The upper envelope -- following closely the upper edge of the signal, thus defining the baseline $B$ -- may simply be obtained by taking the pointwise minimum:
\begin{linenomath}\begin{equation}
B_i\equiv\mathrm{min}\left[M_i^{(\mathrm{forward})},M_i^{(\mathrm{backward})}\right]
\end{equation}\end{linenomath}
Figure \ref{fig5} illustrates the proof of the concept on artificially constructed signals. The straightforward implementation of this procedure is again of $\bigo(N\times P)$ computational complexity. Therefore, a very elegant and efficient algorithm was adopted from Ref.~\cite{max}, that significantly speeds up the procedure, bringing it much closer to the linear dependence: $\bigo(P)$. A simplified version of the code from Ref.~\cite{max}, excluding the calculation of the moving minimum and not requiring the \emph{deque} data structure available in C++, is presented in Table~\ref{tab3} from \mbox{\ref{code}}. Thus obtained envelope may be additionally tightened in order to obtain a smoother and somewhat less artificial baseline. The tightening code, which is more efficient than a quadratic one, is given in Table~\ref{tab4} from \mbox{\ref{code}}. Figure~\ref{fig6} shows the result of this procedure on a selected portion of a real signal from a gaseous $^3$He detector.


\subsection{$\gamma$-flash removal}

\begin{figure}[t!]
\includegraphics[width=1.\linewidth,keepaspectratio]{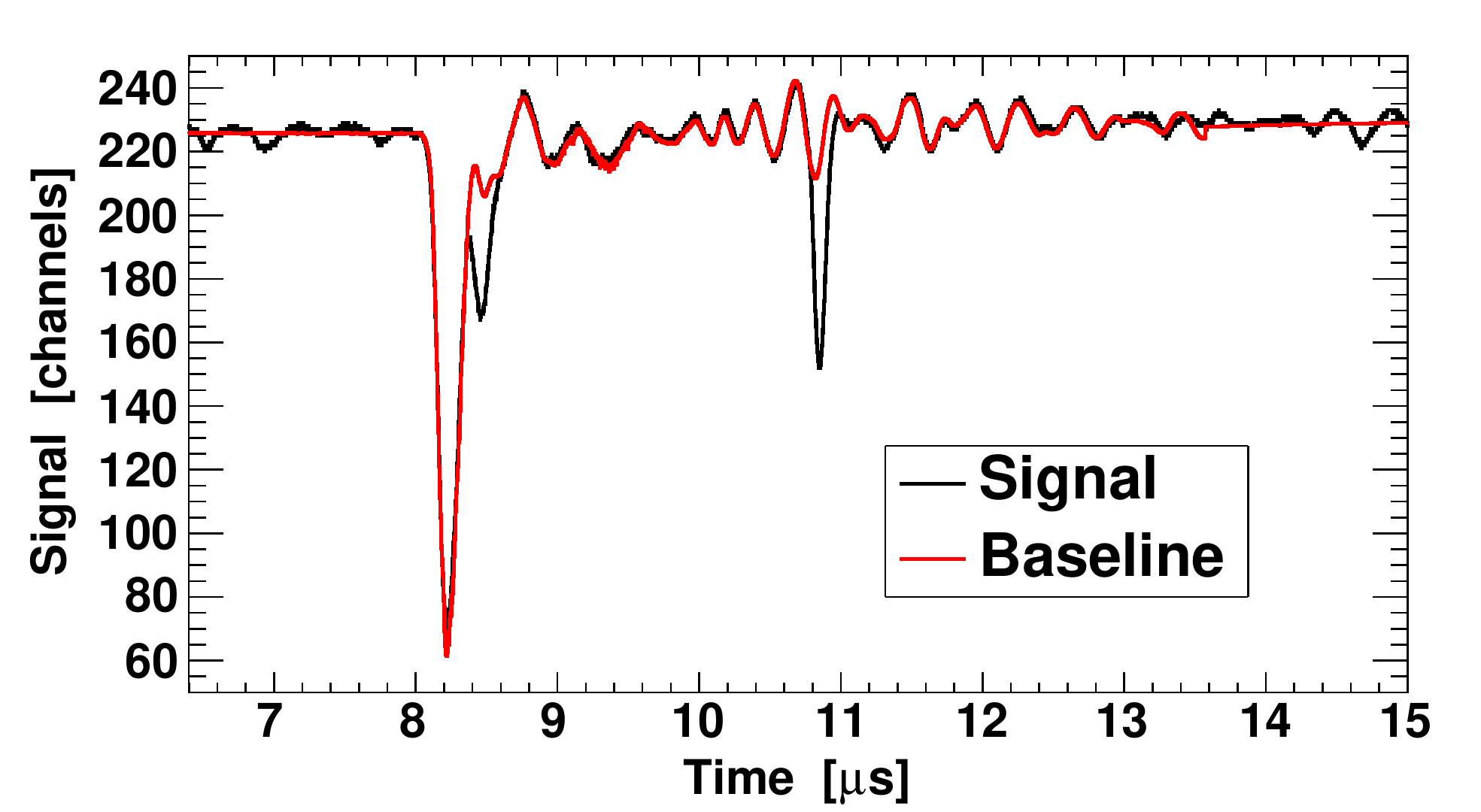}
\caption{(Color online) Adjustment of a distorted baseline to a signal from a MGAS detector. The horizontal adjustment is performed relative to the initial, $\gamma$-flash pulse. The primary (vertical) offset is identified by the constant baseline procedure.}
\label{fig9}
\end{figure}

At neutron time-of-flight facilities the most common cause for a baseline distortion is the induction of a strong pulse by an intense $\gamma$-flash, which is released each time the proton beam hits the spallation target. The response of certain detectors to the $\gamma$-flash is remarkably consistent, which allows for a clear identification of the distorted baseline. By properly averaging a multitude of signals from an immediate vicinity of the $\gamma$-flash pulse, the detector response to a $\gamma$-flash may be recovered in form of an average baseline distortion pulse shape \cite{pu242}. In effect, this pulse shape serves as \emph{a priori} knowledge of the baseline. In general, the baseline offset may be changed for various reasons, e.g. by simply adjusting the digitizer settings. Hence, if available, the shape of the distorted baseline is subtracted from the signal only after identifying and subtracting the primary baseline, which is -- for obvious reasons -- best found as the constant baseline offset. The positioning of the distorted baseline within the signal is performed relative to the $\gamma$-flash pulse, by fitting the externally selected portion of the pulse shape to a leading edge of the $\gamma$-flash pulse. The fitting routine, which is the same as for the regular pulses, is described in Section~\ref{fitting}. Figure~\ref{fig9} shows an example of the adjustment of a distorted baseline to a signal from a MicroMegas detector, clearly revealing the true pulses rising above the baseline, thus providing access to the low time-of-flight, i.e. the high-neutron-energy region.

\section{Pulse shape analysis}
\label{fitting}

After baseline subtraction, the amplitude, area, status of the pileup and timing properties such as the time of arrival are determined for each pulse. Three different methods are available for finding the amplitudes: search for the highest point, parabolic fitting to the top of the pulse and a  predefined pulse shape adjustment. By \emph{pulse shape} we refer to the template pulse of a fixed form, given by the tabulated set of points $(t_i,p_i)$, with $t_i$ as the time coordinate of the $i$-th point and $p_i$ as its height (i.e. the pulse shape value). The optimal pulse shape is best obtained by averaging a large number of real pulses. Several example procedures for excluding unreliable pulses from the pulse shape extraction may be found in Ref.~\cite{psa_sili}.

Though the  pulse shape fitting is generally the most appropriate method for pulse reconstruction, it may not always be applicable, especially if the detector exhibits pulses of strongly varying shapes. This is often the case with gaseous detectors, when the shape and length of the pulse depend on the initial point of ionization and/or the details of the particle trajectory inside the gas volume. The area under the pulse may be calculated by simple signal  integration or from a pulse shape fit, if the latter option has been activated by means of the external input parameter. Finally, extraction of the timing properties relies on the digital implementation of the constant fraction discrimination, with a constant fraction factor of 30\%.

\subsection{Pulse shape fitting -- the numerical procedure}

Pulse shape fitting is a well established method \cite{psa_sili,psa_scint,psa_review}. However, its straightforward implementation is of  $\bigo(n^2)$ computational complexity -- with $n$ as the number of points comprising a typical pulse -- whereas our adopted procedure requires only $\bigo(n\log n)$ operations per pulse. It is important to note that any pulse shape from the following procedure is of the same sampling rate as the analyzed signal. If there is an initial mismatch between the sampling rates of the externally delivered pulse shape and the real signal, the pulse shape is first synchronized to the signal by means of linear interpolation.

Let us consider the predefined (and already synchronized) pulse shape $p$, consisting of $N$ points, with the $M$-th one as the highest point (\mbox{$0\le M\le N-1$}). For a given pulse within the analyzed signal, the left and right fitting boundaries $L$ and $R$ are determined. These may correspond to the pulse boundaries coming directly from the pulse recognition procedure or may be further modified, depending on the pulse requirements. The pulse shape is shifted along the pulse, so that at each step the $M$-th pulse shape point is aligned with an $i$-th pulse point, where $i$ is confined by the fitting boundaries: $i\in[L,R]$. At every position the least squares optimization is performed by minimizing the sum or residuals:
\begin{linenomath}\begin{equation}
\label{eq7}
\mathcal{R}_i\equiv \sum_{j=\max[L\:,i-N_1]}^{\min[R,i+N_2]}\hspace{-5mm}(s_j-\alpha_i p_{j-i+N_1})^2
\end{equation}\end{linenomath}
where by $N_1$ and $N_2$ we have introduced the number of pulse shape points at each side of the $M$-th one:
\begin{linenomath}\begin{align}
\begin{split}
N_1&=M\\
N_2&=N-1-M
\end{split}
\end{align}\end{linenomath}
At each alignment position an optimal multiplicative factor $\alpha_i$ is found from the minimization requirement: $\partial\mathcal{R}_i/\partial\alpha_i=0$. Introducing the following terms:
\begin{linenomath}\begin{align}
\label{eq4}
\begin{split}
S_i&\equiv\sum_{j=\max[L\:,i-N_1]}^{\min[R,i+N_2]}\hspace{-5mm}s_j^2\\
P_i&\equiv\sum_{j=\max[L\:,i-N_1]}^{\min[R,i+N_2]}\hspace{-5mm}p_{j-i+N_1}^2\\
C_i&\equiv\sum_{j=\max[L\:,i-N_1]}^{\min[R,i+N_2]}\hspace{-5mm}s_jp_{j-i+N_1}
\end{split}
\end{align}\end{linenomath}
the optimal $\alpha_i$ may be expressed as:
\begin{linenomath}\begin{equation}
\alpha_i=\frac{C_i}{P_i}
\end{equation}\end{linenomath}
The quality of the fit is evaluated at each alignment point by means of a reduced $\chi^2$:
\begin{linenomath}\begin{align}
\begin{split}
\chi_i^2&\propto\frac{\mathcal{R}_i}{\left(\min[R,i+N_2]-\max[L\:,i-N_1]+1\right)-2}\\
&=\frac{S_i-C_i^2/P_i}{\min[R,i+N_2]-\max[L\:,i-N_1]-1}
\end{split}
\end{align}\end{linenomath}
where the number of points taken by the fit is reduced by 2 due to 2 degrees of freedom: the horizontal and vertical alignment. A fit with a minimal reduced $\chi^2$ is taken as the best result. 

\begin{table}[b!]
\caption{List of recursive relations for calculating the sums from Eq.~(\ref{eq4}). Different cases cover all possible combinations of summation bounds.}
\label{tab5}
\centering
\begin{tabular}{l}
\hline\hline
\textbf{Pulse shape is contained within the pulse}\\
$\boldsymbol{i-N_1> L\qquad\et\qquad i+N_2\le R}$\\
\hline
$S_i=S_{i-1}-s_{i-1-N_1}^2+s_{i+N_2}^2$\\
$P_i=P_{i-1}$\\
\hline
\textbf{Pulse shape protrudes at the beginning of the pulse}\\
$\boldsymbol{i-N_1\le L\qquad\et\qquad i+N_2\le R}$\\
\hline
$S_i=S_{i-1}+s_{i+N_2}^2$\\
$P_i=P_{i-1}+p_{L-i+N_1}^2$\\
\hline
\textbf{Pulse shape protrudes at the end of the pulse}\\
$\boldsymbol{i-N_1>L\qquad\et\qquad i+N_2> R}$\\
\hline
$S_i=S_{i-1}-s_{i-1-N_1}^2$\\
$P_i=P_{i-1}-p_{R-i+1+N_1}^2$\\
\hline
\textbf{Pulse shape protrudes at both ends of the pulse}\\
$\boldsymbol{i-N_1\le L\qquad\et\qquad i+N_2>R}$\\
\hline
$S_i=S_{i-1}$\\
$P_i=P_{i-1}+p_{L-i+N_1}^2-p_{R-i+1+N_1}^2$\\
\hline\hline
\end{tabular}
\end{table}

Equation~(\ref{eq4}) reveals $\bigo(n^2)$ nature of the procedure, with $n$ typically $n\approx R-L$. However, recursive relations for the terms $S_i$ and $P_i$ may be obtained, allowing for their calculation using only $\bigo(n)$ operations. These relations are listed in Table~\ref{tab5}, according to the manner in which the pulse shape and the fitted portion of the pulse are overlapped. By defining the term-wise inverted array $\tilde{p}$ as $\tilde{p}_i=p_{(N-1)-i}$, it becomes evident that the final $C_i$ term from Eq.~(\ref{eq4}) formally corresponds to a convolution of the partial signal $s$ and a pulse shape $p$. In order to calculate $C_i$ at each alignment point in a least number of operations possible, a Fast Fourier Transform algorithm -- of $\bigo(n\log n)$ computational complexity -- was adopted directly from Ref.~\cite{numc}.

Once the best pulse shape alignment has been found by means of a minimal reduced $\chi^2$, the pulse shape is resampled by linear interpolation, constructing the set of $2K$ intermediate pulse shapes $p^{(k)}$ ($k=\pm1,\ldots,\pm K$). In symbolic and self-evident notation, these intermediate terms may be defined as:
\begin{linenomath}\begin{equation}
p_i^{(k)}=p_{i+\frac{k}{K+1}}
\end{equation}\end{linenomath}
Evidently, one may treat the initial pulse shape $p$ as the \mbox{$(2K+1)$-th} member $p^{(0)}$, allowing to establish the uninterrupted indexing by $k\in[-K,K]$. For intermediate pulse shapes the least squares adjustment by minimization of the associated Eq.~(\ref{eq7}) is performed only at the point of the best alignment of the initial pulse shape $p^{(0)}$, calculating the associated members from Eq.~(\ref{eq4}) by direct summation. The adjustment producing a minimal reduced $\chi^2$ (for any $k\in[-K,K]$) is kept as the final result. The value $K=4$ has been adopted for the PSA framework described in this work.

\subsection{Pulse shape fitting -- the saturated pulses}

An important feature of the adopted pulse shape fitting routines is the exclusion of saturated points from the fitting procedure. Here, saturation is defined by the recorded signal reaching the boundaries of the data range (i.e. the minimal or  maximal channel) supported by the data acquisition system (example in Fig.~\ref{fig1}). The saturation management may be directly implemented in Eq.~(\ref{eq7}) through the introduction of appropriate weighting factors $\theta_i$, taking the values 0 or 1:
\begin{linenomath}\begin{equation}
\label{eq8}
\mathcal{R}_i\equiv \sum_{j=\max[L\:,i-N_1]}^{\min[R,i+N_2]}\hspace{-5mm}\theta_j\times(s_j-\alpha_i p_{j-i+N_1})^2
\end{equation}\end{linenomath}
The weighting factors are given as \mbox{$\theta_i=\mathrm{\Theta}(s_i;\:s_\mathrm{min},s_\mathrm{max})$}, where we have introduced the following useful function:
\begin{linenomath}\begin{equation}
\label{eq9}
\mathrm{\Theta}(s_i;\:s_\mathrm{min},s_\mathrm{max})\equiv\left\{\begin{array}{cc}
1&\mathrm{if}\;\; s_\mathrm{min}<s_i<s_\mathrm{max}\\
0& \mathrm{otherwise}
\end{array}\right.
\end{equation}\end{linenomath}
Following the same procedure as for obtaining the expressions from Eq.~(\ref{eq4}), one arrives at the following generalized terms:
\begin{linenomath}\begin{align}
\label{eq10}
\begin{split}
S_i&\equiv\sum_{j=\max[L\:,i-N_1]}^{\min[R,i+N_2]}\hspace{-5mm}\theta_js_j^2\\
P_i&\equiv\sum_{j=\max[L\:,i-N_1]}^{\min[R,i+N_2]}\hspace{-5mm}\theta_jp_{j-i+N_1}^2\\
C_i&\equiv\sum_{j=\max[L\:,i-N_1]}^{\min[R,i+N_2]}\hspace{-5mm}\theta_js_jp_{j-i+N_1}
\end{split}
\end{align}\end{linenomath}
and to the corresponding expression for the reduced $\chi^2$:
\begin{linenomath}\begin{equation}
\chi_i^2\propto\frac{\mathcal{R}_i}{\displaystyle \sum_{j=\max[L\:,i-N_1]}^{\min[R,i+N_2]}\hspace{-5mm}\theta_j-2}
\end{equation}\end{linenomath}
The drawback of this generalization is immediately evident: the $P_i$ term from Eq.~(\ref{eq10}) has become a convolution, in the same way as the $C_i$ term, thus requiring the application of a Fast Fourier Transform, as opposed to the less computationally expensive recursive relations from Table~\ref{tab5} (recursive relations completely analogous to those from Table~\ref{tab5} may now be used only for the $S_i$ term). Furthermore, under the assumption of properly set parameters of the data acquisition system, the saturated pulses are expected to appear only very rarely. For this reason it is advisable to keep the separate approaches -- the one from Eq.~(\ref{eq7}) for unsaturated pulses and the one from Eq.~(\ref{eq8}) for saturated pulses -- instead of applying the generalized and more computationally expensive procedure to both types of pulses.

\subsection{Pulse shape fitting -- the quality control}

\begin{figure}[t!]
\begin{overpic}
[width=1.0\linewidth,keepaspectratio]{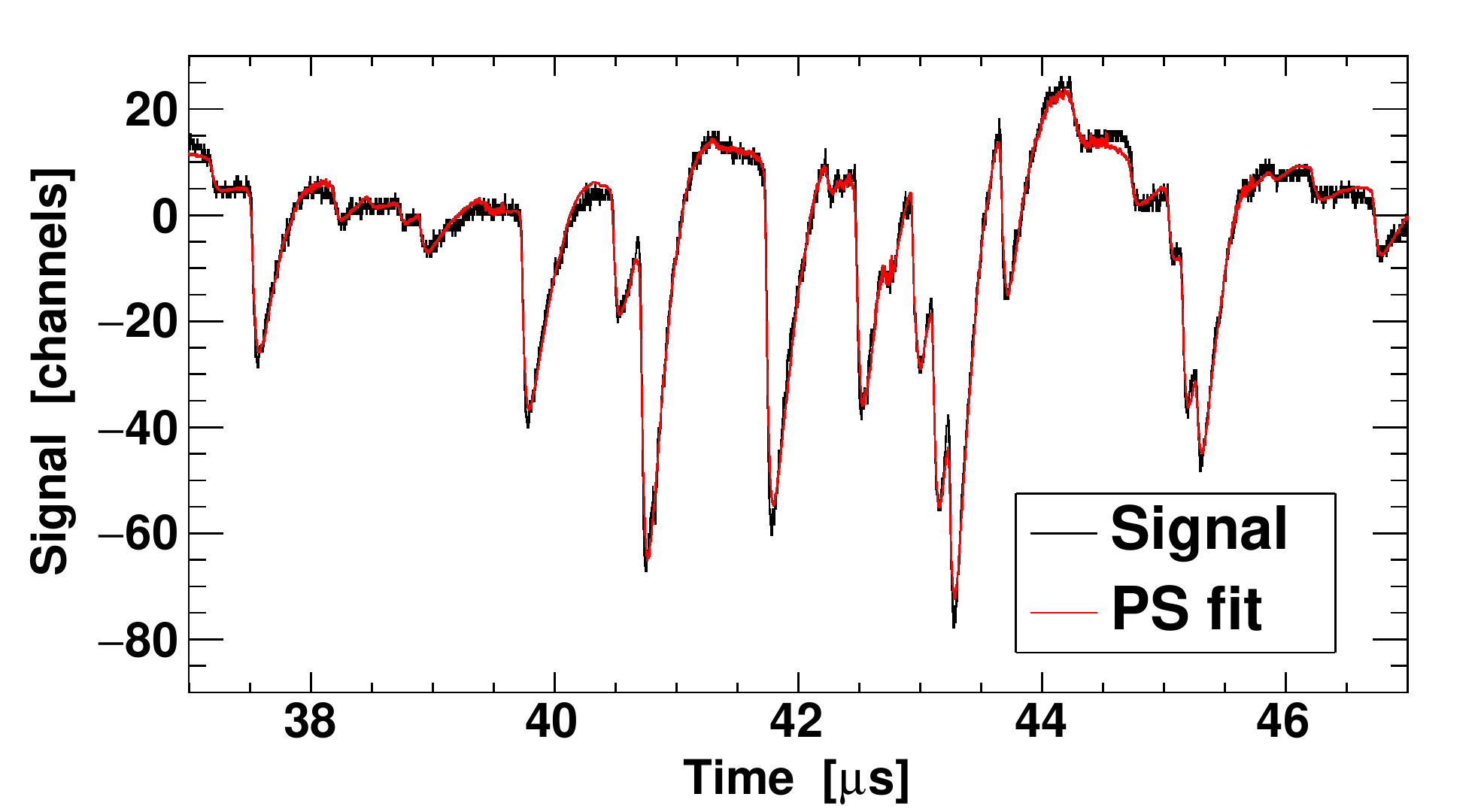}
\put(16.7,10.6){\includegraphics[width=0.18\linewidth,keepaspectratio]{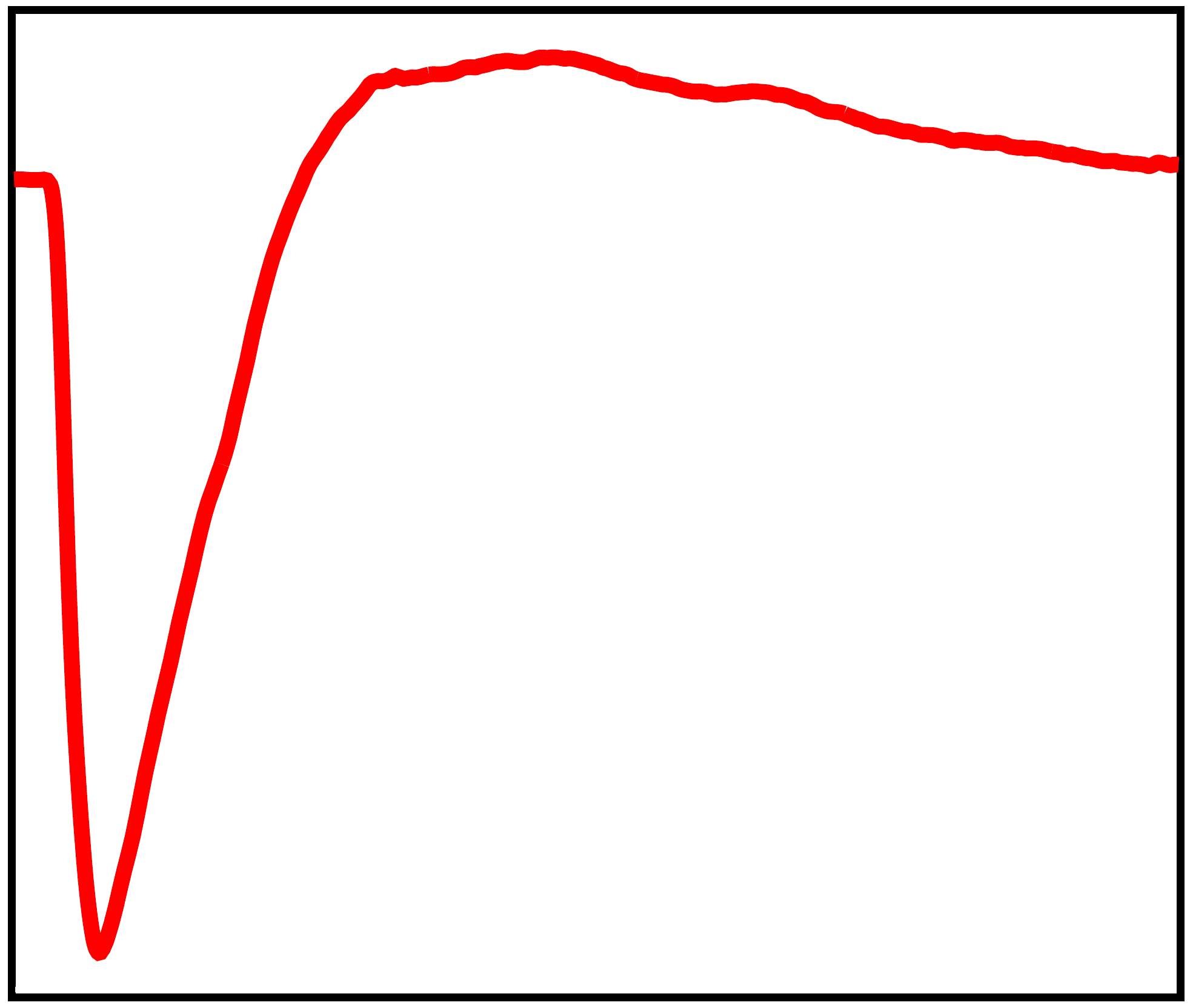}}
\end{overpic}
\caption{(Color online) Signal from a NaI detector characterized by a high density of piled-up pulses. The signal reconstructed by means of pulse shape fitting consists of the fitted and superimposed pulse shapes. Inset shows one of the three pulse shapes used, each adjusted to a given amplitude range.}
\label{fig7}
\end{figure}

Multiple pulse shapes may be provided as input to the program. In this case the pulse shape adjustment is performed for each pulse shape separately and among all fits, the one with the minimal reduced $\chi^2$ is kept. Allowing for the intake of multiple pulse shapes is not only beneficial to detectors exhibiting considerably differing pulses, but was also found specially suitable when the shape of the pulse varies slightly with its amplitude. Hence, among multiple pulse shapes that may be delivered, each may be best suited to a certain amplitude range. In addition, after each adjustment a fitted pulse shape is subtracted from the signal before proceeding to the next pulse in line. Thus, the pulse shape fitting is fully able to account and correct for pileup effects. Figure~\ref{fig7} shows an example of a demanding signal from a NaI detector -- exhibiting a persistent pileup of bipolar pulses -- and a complete signal reconstruction by means of pulse shape fitting. Three separate pulse shapes were used, each adjusted to a given amplitude range. One is shown in an inset of Fig.~\ref{fig7}.

An additional pulse shape fitting control was implemented in form of \emph{discrepancy} -- a quantity similar to the reduced $\chi^2$. Let the fitted pulse shape $f$ be aligned with the pulse in the original signal $s$, so that the index-to-index correlation $s_i\leftrightarrow f_i$ is established (we remind that the optimal pulse shape alignment is determined during the fitting procedure). For the total of $Q$ pulses, let $\alpha_q$ and $\beta_q$ be the indices of the first and the last point of the $q$-th pulse ($q\in[1,Q]$) in the signal. Similarly, let $\mathrm{A}_q$ and $\mathrm{B}_q$ be the first and the last index of the pulse shape aligned to the $q$-th pulse. The discrepancy $D_q$ for the $q$-th pulse is calculated taking into account all the pulse shape points around the fitted pulse -- even if they are outside the fitting range -- as long as the pulse shape does not intrude into any of the neighboring pulses. In addition, the fitted pulse shape point $f_i$ is taken into account if and only if it is between the signal saturation boundaries $s_\mathrm{min}$ and $s_\mathrm{max}$, even if the signal $s_i$ itself is saturated. An explicit expression for the discrepancy $D_q$ takes the form:
\begin{linenomath}\begin{equation}
\label{eq6}
D_q^2\equiv\frac{\displaystyle\sum_{i=\mathrm{max}[\mathrm{A}_q,\:\beta_{q-1}+1]}^{\mathrm{min}[\mathrm{B}_q,\:\alpha_{q+1}-1]}\hspace*{-5mm}\mathrm{\Theta}(f_i;\:s_\mathrm{min},s_\mathrm{max})\times(s_i-f_i)^2}{\displaystyle h^2\times\hspace*{-5mm}\sum_{i=\mathrm{max}[\mathrm{A}_q,\:\beta_{q-1}+1]}^{\mathrm{min}[\mathrm{B}_q,\:\alpha_{q+1}-1]}\hspace*{-5mm}\mathrm{\Theta}(f_i;\:s_\mathrm{min},s_\mathrm{max})}
\end{equation}\end{linenomath}
with $\beta_0=-1$ and $\alpha_{Q+1}=P$ (where $P$ is the total number of points comprising the signal $s$). The $\Theta$-function is defined by Eq.~(\ref{eq9}) (note $f_i$ in place of the first argument). If the discrepancy exceeds the preset threshold value, which is set as an external input parameter, the fit is rejected.

Central to the scaling of $D_q$ is the pulse height $h$ determined directly from the highest point of the baseline-corrected signal; not from the height of the fitted pulse shape. As opposed to $\chi^2$, the discrepancy has the following advantages:
\begin{itemize}\itemsep0pt
\item Due to the pulse height $h$ replacing the signal baseline RMS, the high pulses -- which are well discriminated from the baseline -- are clearly favored by the lower discrepancy values, while the fits to the lower pulses are more susceptible to rejection.
\item In case of any systematic difference between the given pulse shape and the pulses in the signal, the terms $s_i-f_i$ from Eq.~(\ref{eq6}) scale with the pulse height $h$; scaling the discrepancy by the same factor compensates for this effect, canceling the negative bias towards the higher pulses.
\end{itemize}
In addition, adopting the condition expressed through the $\mathrm{\Theta}(f_i;\:s_\mathrm{min},s_\mathrm{max})$ term helps in rejecting the exaggerated fits to severely saturated pulses, such as the ones caused by an intense $\gamma$-flash. When such pulse is saturated for a longer time than a regular pulse would be, only the steep leading edge of the pulse is fitted, due to the exclusion of the saturated points. By rejecting these fits, a subtraction of the overscaled pulse tails is avoided during the pileup correction procedure.

\begin{figure}[t!]
\includegraphics[width=1.\linewidth,keepaspectratio]{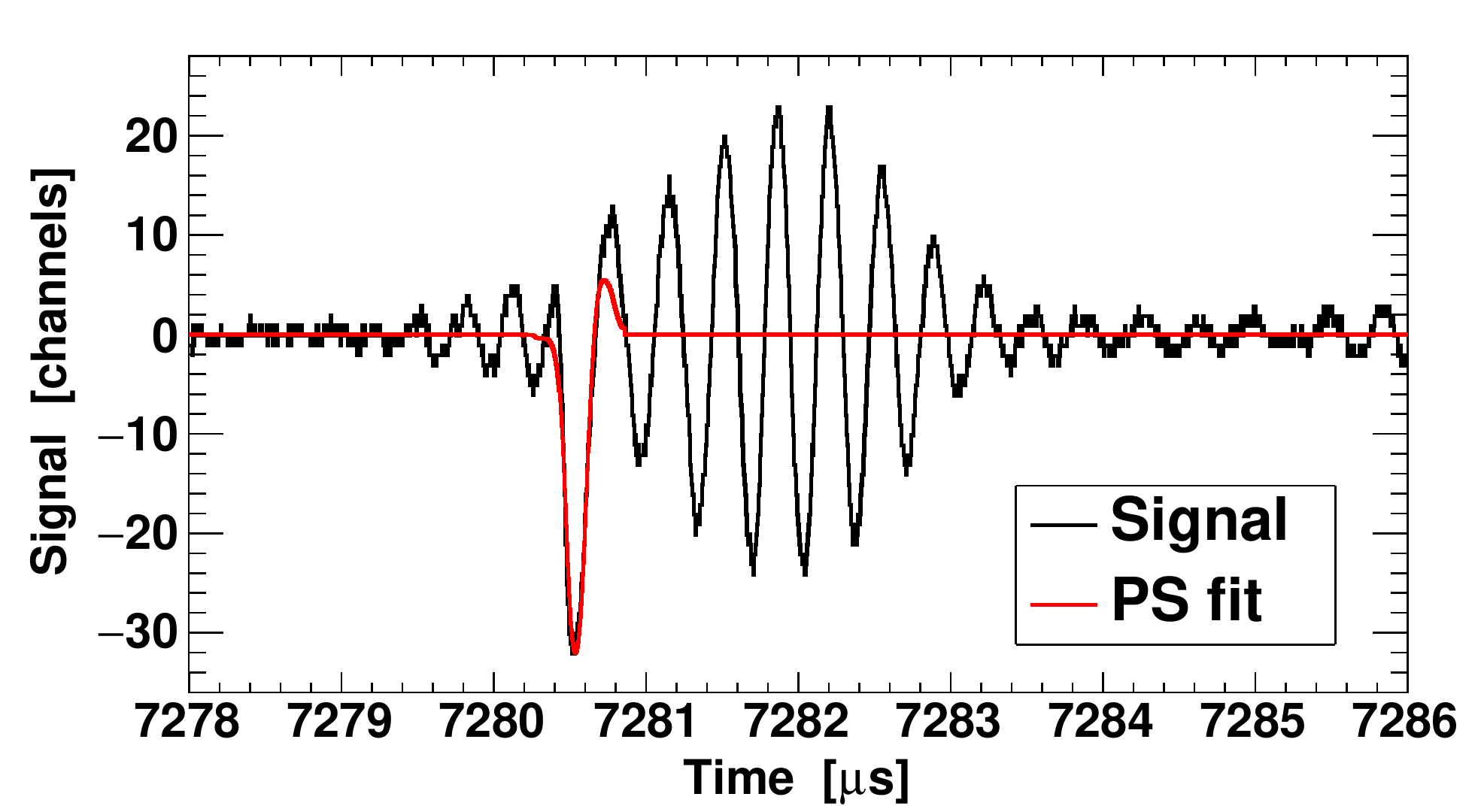}
\caption{(Color online) Example of the pulse rejection capabilities, based only on the calculated discrepancy between the signal and the adjusted pulse shape.}
\label{fig8}
\end{figure}

Figure~\ref{fig8} shows an example of the powerful pulse rejection capabilities, based only on the properly set discrepancy threshold. The single fitted pulse is clearly meaningful, since it significantly deviates from the envelope of the noise. Initially, each of the signal oscillations within a beat is recognized as a potential pulse. Since the shape of these false pulses is incompatible with the given pulse shape, the calculated discrepancy is large and the fit is rejected.

\section{Conclusions}
\label{conclusion}

The most prominent features of the new pulse shape analysis framework developed for the n\_TOF-Phase3 have been described, including the pulse recognition, the baseline calculation and the pulse shape fitting procedures. The pulse recognition relies on the calculation of a custom derivative, as a difference between the signal integrals from both sides of a given point. A supporting procedure for defining the derivative crossing threshold was also described, which isolates the approximate root mean square of the derivative baseline, effectively rejecting the contribution from the beats and actual pulses, while avoiding the dependence on the well defined number of clear presamples.

Three different baseline calculation procedures have been adopted. The simplest one is the constant baseline, which requires a single pass through a signal, without any need for iterative techniques. One of two adaptive baseline options relies on the weighted averaging of the signal, being appropriate when clear portions of the baseline are indeed at hand. The second option is appropriate when this condition is not met -- due to persistent pileup of pulses, completely concealing the baseline -- and no \emph{a priori} knowledge about the baseline is available. In this case the baseline is found as the upper signal envelope, since all regular pulses are treated as negative. In case some \emph{a priori} knowledge of the baseline is available -- coming from a consistent detector response to an intense $\gamma$-flash -- the baseline distortion may be identified in a form of an appropriate pulse shape and may be subtracted from the signal, but only after correcting for the primary baseline offset.

The most basic implementations of previous procedures are of $\bigo(N\times P)$ computational complexity, with $P$ as the total number of points in a digitized signal waveform and $N$ as a characteristic filter width of arbitrary size. Single waveforms recorded by the digital data acquisition system at n\_TOF may, at present, reach the order of magnitude of $10^8$ points. Hence, the $\bigo(N\times P)$ complexity constitutes a significant performance issue that had no alternative but to be resolved. Therefore, for all such procedures fast recursive algorithms were implemented, bringing the computational complexity to the $\bigo(P)$, or at least to the approximate $\bigo(P)$ level. For the reasons of computational efficiency the pulse shape fitting routine was also described, though the procedure itself is well established. By the virtue of a complete \emph{a priori} knowledge of the pulses, the pulse shape fitting procedure allows to subtract the adjusted pulse shapes from the signal, thus correcting for pileup effects and restoring both the energy and timing resolution of the detectors which are considerably affected by pileup.

\section*{Acknowledgments}

This work was supported by the Croatian Science Foundation under Project No. 1680.

\appendix

\section{Moving maximum code}
\label{code}

\begin{table*}[t!]
\caption{Simplified version of the code from Ref.~\cite{max}, adopted for the calculation of the moving maximum. Code input consists of the array \texttt{signal} and the integer parameters \texttt{N}, \texttt{start\_at} and \texttt{stop\_at}. Arrays \texttt{max}, \texttt{max\_forwards} and \texttt{max\_backwards} are to be initialized in advance, having the same number of points as the array \texttt{signal}. At the end of the procedure, array \texttt{max} holds the signal envelope as the final result.}
\label{tab3}
\centering
\begin{tabular}{l|l|l}
\hline\hline
&\textbf{Forward maximum}&\textbf{Backward maximum}\\
\hline\hline
\texttt{1}&\multicolumn{2}{c}{\texttt{\#define MIN(a,b) (a<b?a:b)}}\\
\texttt{2}&\multicolumn{2}{c}{\texttt{int i,U\_first,U\_last;}}\\
\texttt{3}&\multicolumn{2}{c}{\texttt{int U[(const int)(stop\_at-start\_at+1)];}}\\
\\
\texttt{4}&\texttt{U\_first=0;}&\texttt{U\_first=0;}\\
\texttt{5}&\texttt{U\_last=0;}&\texttt{U\_last=0;}\\
\texttt{6}&\texttt{U[U\_first]=start\_at;}&\texttt{U[U\_first]=stop\_at;}\\
\texttt{7}&\texttt{for (i=start\_at; i<=stop\_at; i++) \{}&\texttt{for (i=stop\_at; i>=start\_at; i--) \{}\\
\texttt{8}&\texttt{~~if (U[U\_first]==i-N)}&\texttt{~~if (U[U\_first]==i+N)}\\
\texttt{9}&\texttt{~~~~U\_first++;}&\texttt{~~~~U\_first++;}\\
\texttt{10}&\texttt{~~while (U\_last>=U\_first \&\&}&\texttt{~~while (U\_last>=U\_first \&\&}\\
\texttt{11}&\texttt{~~~~~~~~~signal[i]>=signal[U[U\_last]]) }&\texttt{~~~~~~~~~signal[i]>=signal[U[U\_last]]) }\\
\texttt{12}&\texttt{~~~~U\_last--;}&\texttt{~~~~U\_last--;}\\
\texttt{13}&\texttt{~~U[++U\_last]=i;}&\texttt{~~U[++U\_last]=i;}\\
\texttt{14}&\texttt{~~max\_forwards[i]=signal[U[U\_first]];}&\texttt{~~max\_backwards[i]=signal[U[U\_first]];}\\
\texttt{15}&\texttt{\}}&\texttt{\}}\\
\\
\texttt{16}&\multicolumn{2}{c}{\texttt{for (i=start\_at; i<=stop\_at; i++) max[i]=MIN(max\_forwards[i],max\_backwards[i]);}}\\
\hline\hline
\end{tabular}
\end{table*}


\begin{table*}[t!]
\caption{Code for tightening the signal envelope calculated by the code from Table~\ref{tab3}. The final result is again stored in the array \texttt{max}, i.e. its contents are overwritten.}
\label{tab4}
\centering
\begin{tabular}{l|l|l}
\hline\hline
&\textbf{Forward tightening}&\textbf{Backward tightening}\\
\hline\hline
\texttt{1}&\multicolumn{2}{c}{\texttt{\#define MAX(a,b) (a>b?a:b)}}\\
\texttt{2}&\multicolumn{2}{c}{\texttt{int i,j,last,node,NODES;}}\\
\texttt{3}&\multicolumn{2}{c}{\texttt{int index[(const int)(stop\_at-start\_at+1)];}}\\
\texttt{4}&\multicolumn{2}{c}{\texttt{double slope,last\_slope,past\_slope,A,B;}}\\
\\
\texttt{5}&\texttt{index[0]=start\_at;}&\texttt{index[0]=stop\_at;}\\
\texttt{6}&\texttt{NODES=1;}&\texttt{NODES=1;}\\
\texttt{7}&\texttt{last\_slope=-1.e300;}&\texttt{last\_slope=1.e300;}\\
\texttt{8}&\texttt{past\_slope=-1.e300;}&\texttt{past\_slope=1.e300;}\\
\texttt{9}&\texttt{last=start\_at;}&\texttt{last=stop\_at;}\\
\texttt{10}&\texttt{for (i=start\_at+1; i<=stop\_at; i++) \{}&\texttt{for (i=stop\_at-1; i>=start\_at; i--) \{}\\
\texttt{11}&\texttt{~~slope=(max[i]-max[last])/(x[i]-x[last]);}&\texttt{~~slope=(max[i]-max[last])/(x[i]-x[last]);}\\
\texttt{12}&\texttt{~~if (last\_slope>past\_slope \&\&}&\texttt{~~if (last\_slope<past\_slope \&\&}\\
\texttt{13}&\texttt{~~~~~~last\_slope>slope) \{}&\texttt{~~~~~~last\_slope<slope) \{}\\
\texttt{14}&\texttt{~~~~index[NODES++]=i-1;}&\texttt{~~~~index[NODES++]=i+1;}\\
\texttt{15}&\texttt{~~~~last=i-1;}&\texttt{~~~~last=i+1;}\\
\texttt{16}&\texttt{~~~~last\_slope=0;}&\texttt{~~~~last\_slope=0;}\\
\texttt{17}&\texttt{~~~~slope=(max[i]-max[last])/(x[i]-x[last]);}&\texttt{~~~~slope=(max[i]-max[last])/(x[i]-x[last]);}\\
\texttt{18}&\texttt{~~\}}&\texttt{~~\}}\\
\texttt{19}&\texttt{~~past\_slope=last\_slope;}&\texttt{~~past\_slope=last\_slope;}\\
\texttt{20}&\texttt{~~last\_slope=slope;}&\texttt{~~last\_slope=slope;}\\
\texttt{21}&\texttt{\}}&\texttt{\}}\\
\texttt{22}&\texttt{index[NODES++]=stop\_at;}&\texttt{index[NODES++]=start\_at;}\\
&&\\
\texttt{23}&\texttt{last=0;}&\texttt{last=0;}\\
\texttt{24}&\texttt{for (i=1; i<NODES; i++) \{}&\texttt{for (i=1; i<NODES; i++) \{}\\
\texttt{25}&\texttt{~~if (i==last+1) \{}&\texttt{~~if (i==last+1) \{}\\
\texttt{26}&\texttt{~~~~A=(max[index[i]]-max[index[last]])/}&\texttt{~~~~A=(max[index[i]]-max[index[last]])/}\\
\texttt{27}&\texttt{~~~~~~(x[index[i]]-x[index[last]]);}&\texttt{~~~~~~(x[index[i]]-x[index[last]]);}\\
\texttt{28}&\texttt{~~~~node=i;}&\texttt{~~~~node=i;}\\
\texttt{29}&\texttt{~~\} else if (index[i]-index[last]<=N) \{}&\texttt{~~\} else if (index[last]-index[i]<=N) \{}\\
\texttt{30}&\texttt{~~~~slope=(max[index[i]]-max[index[last]])/}&\texttt{~~~~slope=(max[index[i]]-max[index[last]])/}\\
\texttt{31}&\texttt{~~~~~~~~~~(x[index[i]]-x[index[last]]);}&\texttt{~~~~~~~~~~(x[index[i]]-x[index[last]]);}\\
\texttt{32}&\texttt{~~~~if (slope>=A) \{}&\texttt{~~~~if (slope<=A) \{}\\
\texttt{33}&\texttt{~~~~~~A=slope;}&\texttt{~~~~~~A=slope;}\\
\texttt{34}&\texttt{~~~~~~node=i;}&\texttt{~~~~~~node=i;}\\
\texttt{35}&\texttt{~~~~\}}&\texttt{~~~~\}}\\
\texttt{36}&\texttt{~~\}}&\texttt{~~\}}\\
\texttt{37}&\texttt{~~if (index[i]-index[last]>=N || i==NODES-1) \{}&\texttt{~~if (index[last]-index[i]>=N || i==NODES-1) \{}\\
\texttt{38}&\texttt{~~~~B=max[index[last]]-A*x[index[last]];}&\texttt{~~~~B=max[index[last]]-A*x[index[last]];}\\
\texttt{39}&\texttt{~~~~for (j=index[last]; j<=index[node]; j++)}&\texttt{~~~~for (j=index[last]; j>=index[node]; j--)}\\
\texttt{40}&\texttt{~~~~~~max\_forwards[j]=A*x[j]+B;}&\texttt{~~~~~~max\_backwards[j]=A*x[j]+B;}\\
\texttt{41}&\texttt{~~~~last=node;}&\texttt{~~~~last=node;}\\
\texttt{42}&\texttt{~~~~i=last;}&\texttt{~~~~i=last;}\\
\texttt{43}&\texttt{~~\}}&\texttt{~~\}}\\
\texttt{44}&\texttt{\}}&\texttt{\}}\\
\\
\texttt{45}&\multicolumn{2}{c}{\texttt{for (i=start\_at; i<=stop\_at; i++) max[i]=MAX(max\_forwards[i],max\_backwards[i]);}}\\
\hline\hline
\end{tabular}
\end{table*}

Table~\ref{tab3} presents a computationally efficient C++ code for finding the upper envelope of a signal. The code is a simplifed version of the one proposed in Ref.~\cite{max}. The array \texttt{signal} contains the signal. The external parameters \texttt{N}, \texttt{start\_at} and \texttt{stop\_at} define, respectively: the moving window width, the starting point and stopping point ($0\le$\texttt{start\_at}$<$\texttt{stop\_at}$\le P-1$) of the fraction of the waveform to be taken into account. The arrays \texttt{max}, \texttt{max\_forwards} and \texttt{max\_backwards} are of the same length as the array \texttt{signal} (thus establishing one-to-one correspondence between the array terms; if necessary, the code can also be adjusted so as to use only \texttt{stop\_at-start\_at+1} points for the array \texttt{max} and to completely avoid arrays \texttt{max\_forwards} and \texttt{max\_backwards}). At the end of the procedure, the baseline, i.e. the signal envelope is stored in array \texttt{max}.

Table~\ref{tab4} presents the code for tightening the envelope obtained using the procedure from Table~\ref{tab3}. The main inputs to this code are an array \texttt{x} of positions of signal points and an array \texttt{max} from the previous procedure. As before, arrays \texttt{max\_forwards} and \texttt{max\_backwards} are only used as convenient temporary storage. The code proceeds by identifying the nodes, which define the locally steepest lines, when drawn from a previous node. The set of nodes is, in general, different when searched from the beginning or the end of the waveform. It was empirically found that the initialization \texttt{last\_slope=0} from line 16 is specially favorable -- in contrast to initializations to extreme values -- improving the quality of the tightened baseline. The nodes are then checked for a maximum of the slope between them, within a window of a preset width. If no node is contained within this window, the next available node is used. It is to be noted from lines 29 and 37 that the same moving window width \texttt{N} was used for this procedure as for finding the initial (untightened) envelope. From the results obtained by going forwards and backwards through the waveform, a final tightened one is determined as the pointwise maximum between the two.



\begin{thebibliography}{00}
\bibitem{ntof1} C. Rubbia, S. Andriamonje, D. Bouvet-Bensimon, et al., \emph{A high Resolution Spallation driven Facility at the CERN-PS to
measure Neutron Cross Sections in the Interval from 1 eV to 250 MeV}, CERN/LHC/98-02 (1998).
\bibitem{ntof2} C. Rubbia, S. Andriamonje, D. Bouvet-Bensimon, et al., \emph{A high Resolution Spallation driven Facility at the CERN-PS to
measure Neutron Cross Sections in the Interval from 1 eV to 250 MeV}, CERN/LHC/98-02-Add. 1 (1998).
\bibitem{ear2_0} E. Chiaveri, \emph{Proposal for n\_TOF Experimental Area 2 (EAR-2)}, CERN-INTC-2012-029/INTC-O-015 (2012).
\bibitem{ear2_1} C. Wei{\ss}, E. Chiaveri, S. Girod, et al., Nucl. Instr. and Meth. A 799 (2015) 90.
\bibitem{ear2_2} S. Barros, I. Bergstr\"{o}m, V. Vlachoudis and C. Wei\ss, J. Instrum. 10 (2015) P09003.
\bibitem{carlos} C. Guerrero, A. Tsinganis, E. Berthoumieux, et al., Eur. Phys. J. A 49 (2013) 27.
\bibitem{simon} S. Marrone, P. F. Mastinu, U. Abbondanno, et al., Nucl. Instr. and Meth. A 517 (2004) 389.
\bibitem{diamond} C. Wei\ss, E. Griesmayer, C. Guerrero, et al., Nucl. Instr. and Meth. A 732 (2013) 190.
\bibitem{tac} C. Guerrero, U.Abbondanno, G.Aerts, et al., Nucl. Instr. and Meth. A 608 (2009) 424.
\bibitem{c6d6} R. Plag, M. Heil, F. K\"{a}ppeler, et. al., Nucl. Instr. and Meth. A 496 (2003) 425.
\bibitem{mgas1} Y. Giomataris, Ph. Rebourgeard, J.P. Robert and G.Charpak, Nucl. Instr. and Meth. A 376 (1996) 29 .
\bibitem{mgas2} S. Andriamonje, M. Calviani, Y. Kadi, et al., J. Korean Phys. Soc. 59 (2011) 1597.
\bibitem{ptb} D. B. Gayther, Metrologia 27 (1990) 221.
\bibitem{ppac} C. Paradela, L. Tassan-Got, L. Audouin, et al., Phys. Rev. C 82 (2010) 034601.
\bibitem{daq} U. Abbondanno, G. Aerts, F. \'{A}lvarez, et al., Nucl. Instr. and Meth. A 538 (2005) 692.
\bibitem{baf2_analysis} E. Berthoumieux, \emph{Preliminary report on BaF$_2$ Total Absorption Calorimeter test measurement}, Rap. Tech. CEA-Saclay/DAPNIA/SPhN (2004).
\bibitem{psa_naid} W. Xiao, A. T. Farsoni, H. Yang, D. M. Hamby, Nucl. Instr. and Meth. A 769 (2015) 5.
\bibitem{psa_hpge} R. J. Cooper, D. C. Radford, K. Lagergren, et al., Nucl. Instr. and Meth. A 629 (2011) 303.
\bibitem{psa_sili} S. N. Liddick, I. G. Darby, R. K.Grzywacz, Nucl. Instr. and Meth. A 669 (2012) 70.
\bibitem{psa_scint} C. Guerrero, D.Cano-Ott, M.Fern\'{a}ndez-Ord\'{o}\~{n}ez, et al., Nucl. Instr. and Meth. A 597 (2008) 212.
\bibitem{psa_review} J. Kamleitner, S. Coda, S. Gnesin, Ph. Marmillod, Nucl. Instr. and Meth. A 736 (2014) 88.
\bibitem{numc} W. H. Press, S. A. Teukolsky, W. T. Vetterling and B. P. Flannery, \emph{Numerical Recipes in C: The Art of Scientific Computing (Second Edition)}, Cambridge: Cambridge University Press (1992). 
\bibitem{hann} F. J. Harris, Proc. IEEE 66, no. 1 (1978) 51.
\bibitem{competitive} M. Nied\'{z}iwiecki, W. A. Sethares, IEEE Trans. Signal Process. 43, no. 1 (1995) 1.
\bibitem{max} D. Lemire, Nord. J. Comput. 13(4) (2006) 328.
\bibitem{pu242} A. Tsinganis, E. Berthoumieux,C. Guerrero, et al., Nucl. Data Sheets 119 (2014) 58.
\end{thebibliography}
\end{document}